\documentclass[useAMS,usenatbib]{mn2e}
\usepackage{graphicx}
\usepackage{rotating}
\usepackage{longtable}
\usepackage{graphicx}
\usepackage{longtable}
\usepackage{lscape}

\def\msol{M$_\odot$ }

\def\lsim{\mathrel{\rlap{\lower 3pt \hbox{$\sim$}} \raise 2.0pt \hbox{$<$}}}
\def\gsim{\mathrel{\rlap{\lower 3pt \hbox{$\sim$}} \raise 2.0pt \hbox{$>$}}}

  \title[AGNs in late-type Virgo galaxies]{The census of nuclear activity of late-type galaxies in the Virgo cluster}

\author[Decarli et al.]{R. Decarli,$^{1,2}$\thanks{roberto.decarli@mib.infn.it} G. Gavazzi,$^1$ I. Arosio,$^1$
L. Cortese,$^3$ A. Boselli,$^4$  C. Bonfanti,$^1$  M. Colpi$^1$\\
             $^1$ Universit\`a degli Studi di Milano-Bicocca, Piazza delle
             scienze 3, 20126 Milano, Italy\\
             $^2$ Universit\`{a} degli Studi dell'Insubria, via Valleggio 11,
             22100 Como, Italy \\
             $^3$ School of Physics and Astronomy, Cardiff University, The Parade, Cardiff CF24 3YB\\
             $^4$ Laboratoire d'Astrophysique de Marseille, BP8, Traverse du Siphon,
             F-13376 Marseille, France \\
       }
\pagerange{\pageref{firstpage}--\pageref{lastpage}} \pubyear{2007}

\begin{document}
\maketitle

\begin{abstract} The first spectroscopic census of AGNs
associated to late-type galaxies in the Virgo cluster is
carried on by observing 213 out of a complete set of 237
galaxies more massive than $M_{\rm dyn}>10^{8.5}$ \msol. Among them,
77 are classified as AGNs (including 21 transition objects, 47
LINERs and 9 Seyferts), and comprize 32 \% of the late-type galaxies
in Virgo. Due to spectroscopic incompleteness at most 21
AGNs are missed in the survey, so that the fraction would increase
up to 41 \%. Using corollary Near-IR observations, that
enable us to estimate galaxies dynamical masses, it is found that
AGNs are hosted exclusively in massive galaxies, i.e. $M_{\rm
dyn}\gsim 10^{10}$ \msol. Their frequency increases steeply with the
dynamical mass from zero at $M_{\rm dyn}\approx10^{9.5} $ \msol to
virtually 1 at $M_{\rm dyn}>10^{11.5} $ \msol. These frequencies are
consistent with the ones of low luminosity AGNs found in the general
field by the SDSS. Massive galaxies that harbor AGNs commonly show
conspicuous $r$-band star-like nuclear enhancements. Conversely they
often, but not necessarily contain massive bulges. Few well known
AGNs (e.g. M61, M100, NGC4535) are found in massive Sc galaxies with
little or no bulge. The AGN fraction seems to be only marginally
sensitive to galaxy environment. We infer the black hole masses
using the known scaling relations of quiescent black holes. No black
holes lighter than $\sim 10^6$ \msol are found active in our sample.
\end{abstract}


\section{Introduction}

Active Galactic Nuclei (AGNs) can be broadly divided in classes of
decreasing luminosity from quasars, Seyfert to Low-Ionization
Nuclear Emission Regions (LINERs) characterized by narrow emission
lines of relatively low ionization and large [NII]/H$\alpha$ line
ratios. A growing body of evidences, triggered by studies of low
luminosity AGNs, from radio to X-rays (Halderson et al. 2001,
Terashima et al. 2002, Filho et al. 2004, Martini et al. 2006) and
in particular in the optical band, mostly owing to the Sloan Digital
Sky Survey (SDSS; York et al. 2000) is reinforcing the scenario
where LINERs represent the missing link between higher luminosity
AGNs and dormant supermassive black holes. Hao et al. (2005)
estimate that 4-10 \% of all galaxies in the SDSS harbor an AGN;
Kauffmann et al. (2003) estimate that up to 80\% of galaxies of more
than $10^{11}$ \msol harbor a supermassive black hole, either
dormant or active.

The role of the environment in triggering or inhibiting nuclear
activity is still not clear. Dressler et al. (1985) observed an AGN
fraction in the field $\sim 5$ times higher than in clusters.
Kauffmann et al. (2004) found that the fraction of powerful AGNs
(with high [OIII] luminosity, i.e. dominated by Seyferts) in the
SDSS decreases with increasing galaxy environment density. They
interpret this result in connection with the fact that strong AGNs
are hosted in star-forming galaxies that avoid dense environments.
Popesso \& Biviano (2006) found that the AGN fraction decreases with
increasing velocity dispersion of galaxies in groups and clusters,
being higher in dense, low-dispersion groups. On the other hand,
Shen et al. (2007) suggest that the fraction of AGNs in galaxy
groups and in clusters may be the same. Miller et al. (2003) find
that the frequency of low activity AGNs is insensitive to the
environment. Furthermore, Boselli \& Gavazzi (2006) showed that
gravitational interaction between the galaxy and the cluster
potential as a whole does trigger gas infall toward the galaxy center,
and this may feed nuclear activity.

The Virgo Cluster provides a perfect environment for studying the
relations between the nuclear activity and the host galaxy, for at
least four reasons: 1) it's near enough to grant us a highly
detailed knowledge of the host galaxy internal structure and
morphology; the nuclear component is also easily disentangled among
inner galaxy regions; 2) a complete survey of a wide range of
luminosity can be achieved; 3) many galaxies member of the Virgo
Cluster have been studied individually in almost all the
electromagnetic spectrum (e.g., Allard et al. 2006; Garca-Burillo,
et al., 2005; Chy\.{z}y et al., 2006; Boselli et al. 2006; Yoshida
et al. 2004); 4) the environmental effects on the nuclear activity
may be examined through various indicators (i.e. gas
deficiency).

Cot\'e et al. (2006), analyzing
100 Elliptical galaxies in the Virgo cluster surveyed with the ACS
on board the HST by Ferrarese et al. (2006a), find that most (66-82
\%) contain nuclear cusps, mainly in galaxies less massive than
$3\cdot 10^{10}$ \msol. Wehner \& Harris (2006) and Ferrarese et al.
(2006b) found that the masses of these bright cusps follow the same
scale relations as the black hole (BH) masses in giant ellipticals,
occupying the same range as the lower-end extrapolation of the
$M_{\rm BH}$ -- Magnitude correlation. This suggests that a common
mechanism is responsible for the growth of both nuclear stellar
cusps or black holes on these mass scales.

As far as late-type galaxies in the Virgo cluster, instead, no
systematic survey exists providing the census of massive BHs nor of
AGNs of the various species. To fill this gap we undertook a survey
of virtually all spiral galaxies brighter than 15 mag belonging to
the Virgo cluster, probing their activity with diagnostic tools that
are sensitive to the presence of low activity AGNs. In spite of the
poorer statistics, the Virgo spirals appear to have AGN properties
consistent with the ones in the SDSS, i.e. their frequency depends
critically on the mass of the hosts galaxies. By exploiting the
superior resolution obtained at the distance of the Virgo cluster we
show that AGNs develop in large mass galaxies that have significant
stellar nuclei, independently from the existence of conspicuous
bulges.

\section{Sample}

The present analysis is based on a volume limited, complete set of
late-type galaxies belonging to the Virgo cluster. The sample is
obtained by selecting from the Virgo Cluster Catalogue (Binggeli et
al. 1985, 1993) all cluster members ($-1000<Vel<3000$ km/s) with
apparent photographic magnitude brighter than 15.0 mag and Hubble
type between Sa and Im/BCD. A distance of $D$=17.0 Mpc is assumed for
all clouds constituting the Virgo cluster, except W, M ($D$=32 Mpc)
and B ($D$=23 Mpc) (see Gavazzi et al. 1999), equivalent to assuming
an Hubble constant of $H_o=75 ~\rm km ~s^{-1} Mpc^{-1}$. The
resulting sample comprises 237 galaxies brighter than $M_{\rm
pg}\approx-16$ mag.

\section{Data}
\begin{figure}
   \includegraphics[width=0.5\textwidth]{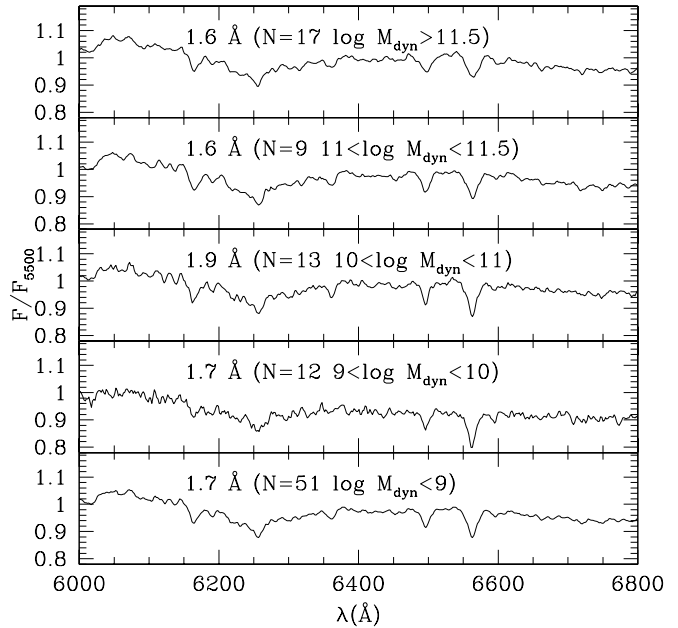}
\caption{\small Red-channel template spectra near H$\alpha$
of early-type galaxies binned in 4 mass intervals and all together (bottom) are used
to estimate the underlying continuum at H$\alpha$. The E.W.
of H$\alpha$, the number of averaged spectra and the mass interval are labeled.
}\label{templ}
\end{figure}

\begin{table}
\caption {Completeness.}
\label{compl}
\begin{center}
$\begin{array}{ccccccccc} \hline \noalign{\smallskip}
m_{\rm pg}\leq15.0        & \textrm{Near-IR}        &   r\textrm{-band} & \rm Spec. classif.     \\
\noalign{\smallskip} \hline \noalign{\smallskip}
237                &    216         &   199    &     213            \\
\noalign{\smallskip}
\hline
\end{array}$
\end{center}
\end{table}
For the aim of the present study, for each of the target galaxies we
collect three sets of observations: Near-IR (H-band) imaging,
optical (Gunn $r$-band) imaging and intermediate resolution (R
$\sim$ 1000) spectroscopy. These quantities are partly derived from
existing observations available from the WEB site GOLDMine (Gavazzi
et al. 2003), partly obtained on purpose for the present
investigation. Table \ref{compl} gives the coverage of the various
data sets, stressing the high degree of completeness of the spectral
classification (213/237=90\%), of the optical (199/237=84\%) and
Near-IR imaging (216/237=91\%). Near-IR imaging was obtained as
described in Gavazzi et al. (2001) (and references therein) and from
2MASS. H-band total luminosities ($L_H$ in L${}_{\odot}$) are
obtained from extrapolation to infinity of light profiles fitted to
isophotal ellipses (Gavazzi et al. 2000). For 21 galaxies without
Near-IR data, $L_H$ is derived from the photographic luminosity:
\begin{equation}\label{magph_hlum}
\log L_H=0.80-0.50(M_{\rm pg})
\end{equation}
where $M_{\rm pg}=m_{\rm pg}-5\log D-25$. Equation \ref{magph_hlum}
was obtained by fitting the H-band luminosity and the $M_{\rm pg}$
of 440 late-type galaxies with available data in the GOLDMine
database. The dispersion of this relation is in the order of $0.2$
dex.

$L_H$ provides us with an estimate of the dynamical mass
within the optical radius of disk galaxies (Gavazzi et al. 1996),
according to $\log M_{\rm dyn}= \log L_H+0.66$. From the Near-IR
images we also derive the H-band light concentration index $C_{31}$,
i.e. the ratio of the radii containing 75 and 25 \% of the total
light. This parameter is sensitive to the presence of conspicuous
central bulges (with $C_{31}$ in excess of 4-5, while pure disks
have $C_{31}\sim 2.8$ -- see Gavazzi et al., 2000).

As part of the H$\alpha$ imaging survey of Virgo spirals, complete
down to $m_{\rm pg}\leq 16.5$ (Gavazzi et al. 2006), $r$(Gunn)
images were obtained to estimate the red stellar continua. Here we
use the 199 images available for the present sample to estimate and
quantify the presence of relevant nuclear enhancements. For each
galaxy we construct the parameter $Nuc(r)$ as the difference of the
$r$-band surface brightness within 1.5 of the seeing disk and the
mean surface brightness between 25 and 50 \% of the $r$-band radius.
Galaxies with $Nuc(r)$ in excess of 1 ($\rm mag ~arcsec^{-2}$) have
significant optical nuclei, harboring approximately $10^9$ years old
stars, or contributed by the AGN continuum radiation\footnote{The
nuclearity parameter was not derived from the Near-IR data-set
because these images are not of sufficient seeing quality to derive
a nuclear parameter, whereas they are adequate for determining
$C_{31}$}.

\begin{table}
\caption {Spectral available data. `Ho' refers to line fluxes published in
Ho, Filippenko \& Sargent (1997); `SDSS' to SDSS DR5 spectra; `Loiano' to Loiano
nuclear spectra; `DS' and `mDS' refer to Drift-Scan and modified Drift-Scan spectra,
as described in the text; `NED' lists the galaxy nuclear classification available in
the NASA/IPAC Extragalactic Database.}
\label{availdata}
\[
%\begin{array}{p{0.5\linewidth}cccccccc}
\begin{array}{ccccccccc}
\hline
\noalign{\smallskip}
%Region &  & \multicolumn{2}{c}{Imaging} & \multicolumn{2}{c}{Spectroscopy}\\
~~~            & \rm Ho           & \rm SDSS  & \rm Loiano  & \rm mDS   & \rm DS & \rm  NED   & \rm  All    \\
\noalign{\smallskip}
\hline
\noalign{\smallskip}
Available      &     40       &   84  &    29 &   22 &    193   &    41  &   409    \\
Adopted        &     40       &   73  &    13 &    4 &     81   &     2  &   213    \\
\noalign{\smallskip}
\hline
\end{array}
\]
\end{table}

\begin{figure*}
\centering
   \includegraphics[width=0.7\textwidth]{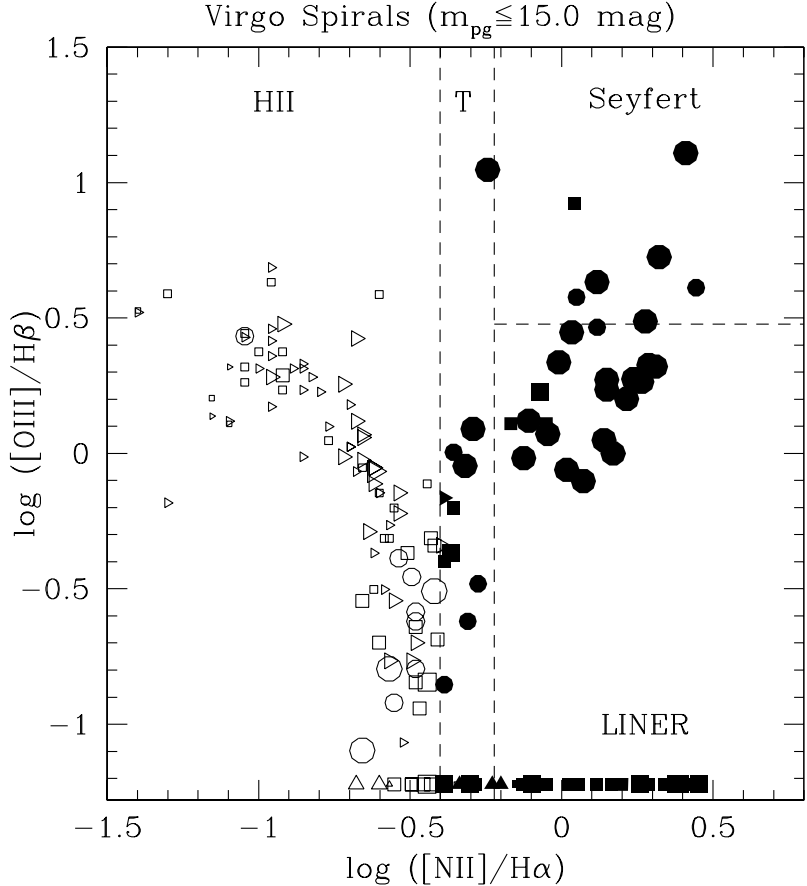}
\caption{\small The 2-D diagnostic diagram (the BPT diagram, from
Baldwin, Phillips \& Terlevich, 1981) used to characterize the
spectra of galaxies in this work based on the [NII]/H$\alpha$ and
[OIII]/H$\beta$ line ratios. Objects for which [OIII]/H$\beta$ is
not available either because the spectra were taken in the
red-channel (Loiano spectra) or lines were too weak to be measured,
are plotted at log[OIII]/H$\beta$ = -1.2. Triangles are for
drift-scan spectra, squares are for the our nuclear spectra
(Loiano+SDSS+modified drift-scan), while circles are from Ho et
al.(1997); empty symbols are for HII-like nuclei, filled symbols for
Transition, LINER and Seyfert. Symbol size is proportional to the
dynamical mass.}\label{diagno}
\end{figure*}

Our analysis is based on the ratios of narrow emission line fluxes.
Different sets of spectroscopic data are used on this purpose. The
first line of Table \ref{availdata} lists all the available
spectroscopic material, indicating that for many of the 237 sampled
galaxies more than one spectroscopic source is available. The second
line gives the 213 independent measurements that were finally
adopted after choosing among various possibilities, according to the
following priorities. We first checked their availability in Ho et
al. (1997). If present we adopt the line ratios, not the final
classification given in that paper. Secondly we measure the line
fluxes on available nuclear spectra. 84 spectra, obtained with
apertures of 3 arcsec, were published in the SDSS in the fifth data
release (Adelman-McCarthy, J., et al. 2007). 73 of these lacked of a
classification in Ho et al. (1997). Other nuclear spectra were
obtained on purpose in February and March 2005, 2006 using the
Bologna-Faint-Object-Spectrograph (BFOSC) attached to the Loiano
1.5m telescope. These consist of nuclear long-slit spectra taken
through a 2-arcsec slit, combined with an intermediate-resolution
grism ($R \sim 1000$) covering however only a limited amount of the
red-channel (6100 - 8200 \AA), containing H$\alpha$ and [NII], but
not H$\beta$ nor [OIII]. Out of the 29 spectra taken at Loiano only
13 were adopted for the final classification because we often
preferred the Ho and SDSS spectra, covering a broader spectral
range.\footnote{Note added in proofs: among the 23 unobserved
galaxies listed at the end of Table \ref{tabellone}, 7 were observed
with the Loiano telescope in spring 2007, leading to the following
results: VCC99, 449, 567, 697, 1442, CGCG14062 (with log$(L_H/\rm
L_{\odot})<9.75$) have  HII-like nuclei. The brightest VCC362
(log$(L_H/\rm L_{\odot})=10.46$) is classified as LINER. The
spectroscopic coverage now reaches 93\% of the  sample.} 193 spectra
available for galaxies in the present study were taken from the
GOLDMine database. These spectra were taken in the drift-scan mode
(labeled ``DS'' after Drift-Scan in Table \ref{availdata}), i.e.
letting the slit of the spectrograph, parallel to the galaxy
major-axis, slide across the minor-axis (Gavazzi et al. 2004).
Spectra taken in this modality maintain their spatial resolution
only along the slit. DS spectra typically cover the 3500-7000 \AA {}
range with $R \sim 600-1000$. Since they are usually employed to
study the overall, luminosity-averaged spectral characteristics of
galaxies, the aperture is kept wide enough to integrate over a large
fraction of the slit.  Twenty-two of the DS spectra were also
re-extracted with a smaller ($\sim 5$ arcsec) aperture. These data
are labeled as ``modified DS''. The role of these data is
further discussed in section \ref{spatial}. Finally, the NASA/IPAC
Extragalactic Database reports, without completeness, the nuclear
classification of 41 objects in our sample. Only 2 of them had no
other spectral data in our study, and were thus classified according
to NED.

In all available spectra (excluding Ho et al. 1997) we measured the
intensity of the [OIII], [NII] and the narrow-component H$\beta$ and
H$\alpha$ emission lines. Underlying H$\beta$ absorption is measured
separately (as explained in Gavazzi et al. 2004) and the adopted
value of H$\beta$ in emission is corrected accordingly. Underlying
H$\alpha$ is not equally easy to measure for most emission-line
objects. To estimate the underlying H$\alpha$ and correct for it one
could use several approaches. The most recommended one would be that
of fitting stellar population synthesis models (e.g. Bruzual \&
Charlot, 2003) to the observed spectra and measuring H$\alpha$ from
the models. The drawback of this method is that it suffers from
uncertainties in the extinction correction, unless UV and Far-IR
data were available, which is not the case for nuclear regions. We
therefore preferred a simpler method based on the assumption that
elliptical-like objects are realistic representations of the
circumnuclear stellar properties of galaxies. We therefore assembled
template spectra of 51 early-type galaxies available from GOLDMine,
binned in 4 mass intervals (each containing tens of spectra). It is
apparent from Figure \ref{templ} that the amount of absorption at
H$\alpha$ is independent of the galaxy luminosity. We therefore
combined all available early-type spectra into a unique template
spectrum that we subtract (after normalization) from the individual
emission line spectra. On average the E.W. of the underlying
H$\alpha$ is $\sim 1.7$ \AA. If line fluxes fell under the
sensitivity limit ($\lsim 3 \sigma$ of the rms of the spectral
continuum), we classified the galaxy as No Emission Line (NEL)
galaxy.

\begin{figure*}
 \includegraphics[width=0.55\textwidth]{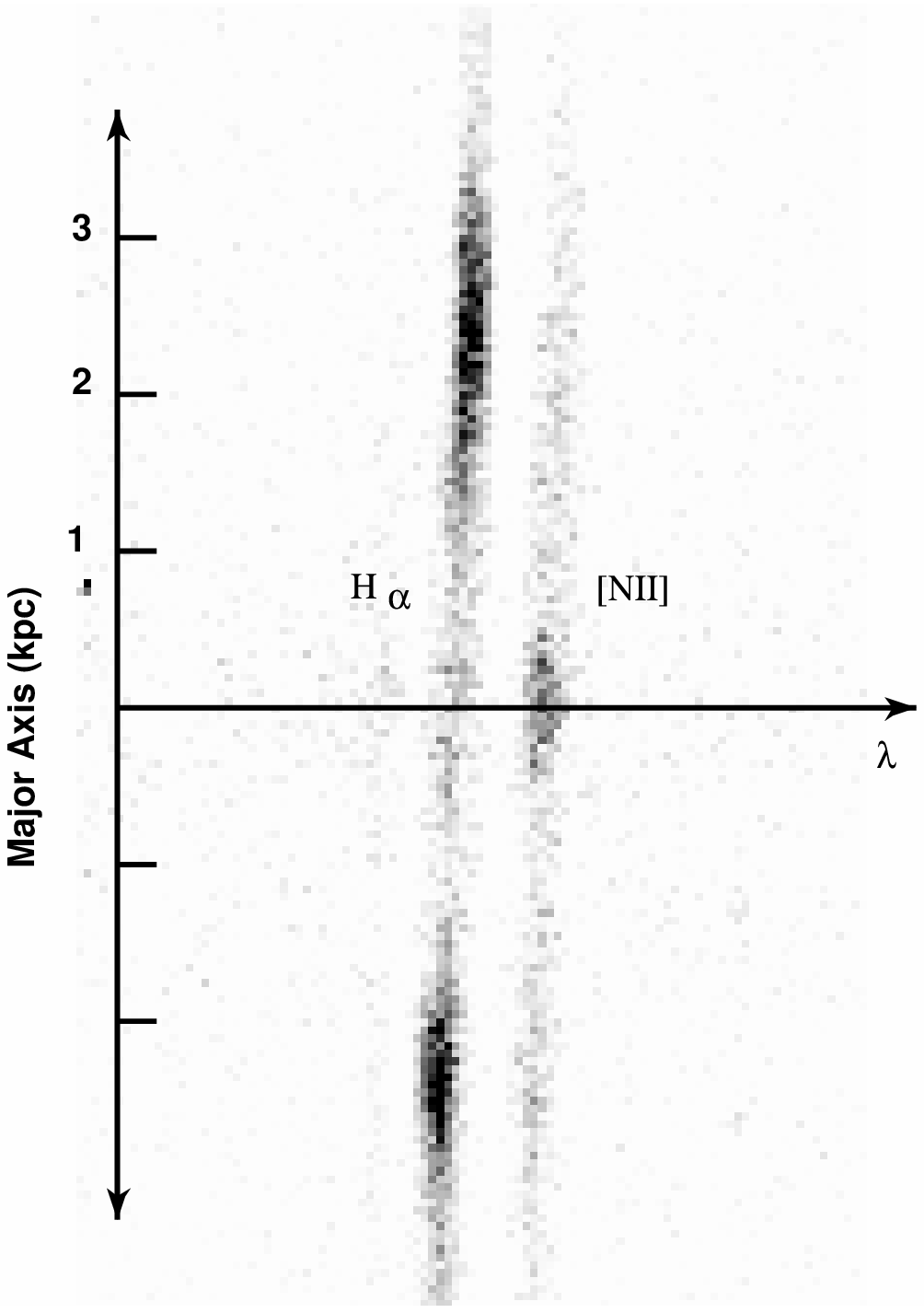}
 \includegraphics[width=0.4\textwidth]{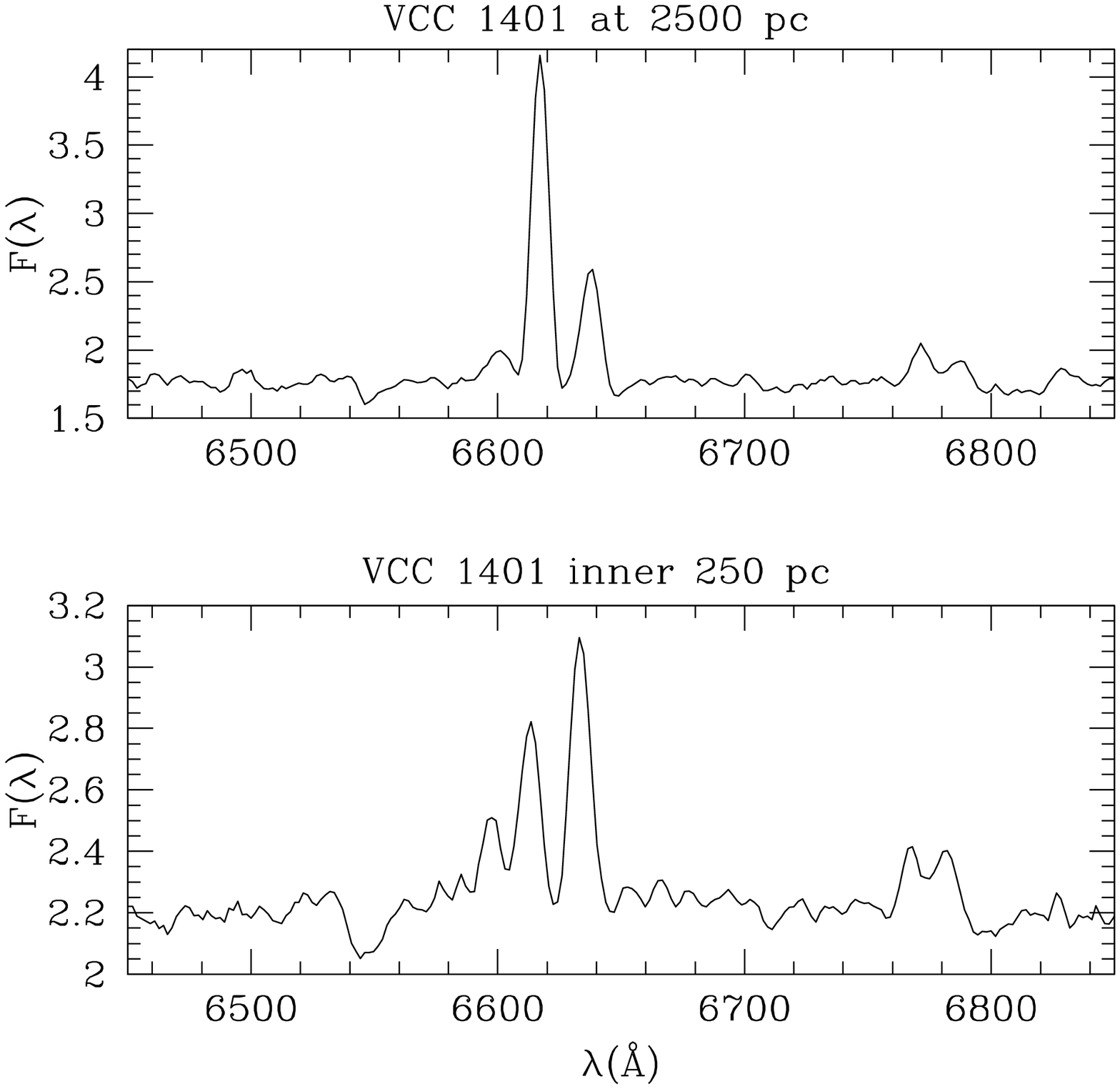}
\caption{\small
Blow-up of the 2-D long-slit Drift-Scan spectrogram of VCC1401 in the region of H$\alpha$, showing
the inversion of the [NII]/H$\alpha$ ratio taking place approximately at 500 pc.
The nuclear spectrum in the inner 250 pc from the center of the AGN VCC1401 (bottom-right panel)
and at 2.5 kpc from the center (top-right panel).}
\label{1401}
\end{figure*}
   \begin{figure*}
   \begin{center}
   \includegraphics[width=0.45\textwidth]{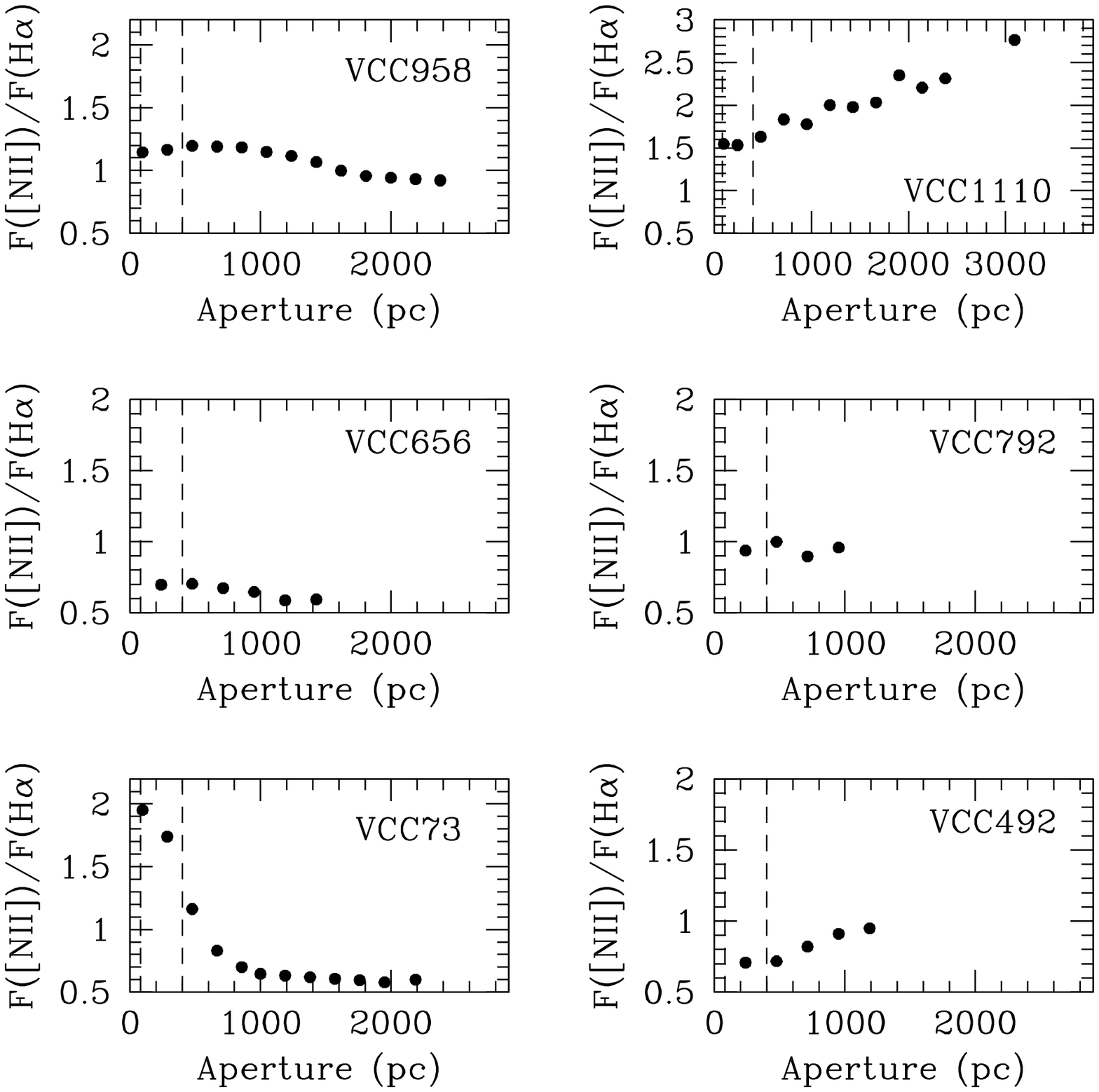}
   \includegraphics[width=0.45\textwidth]{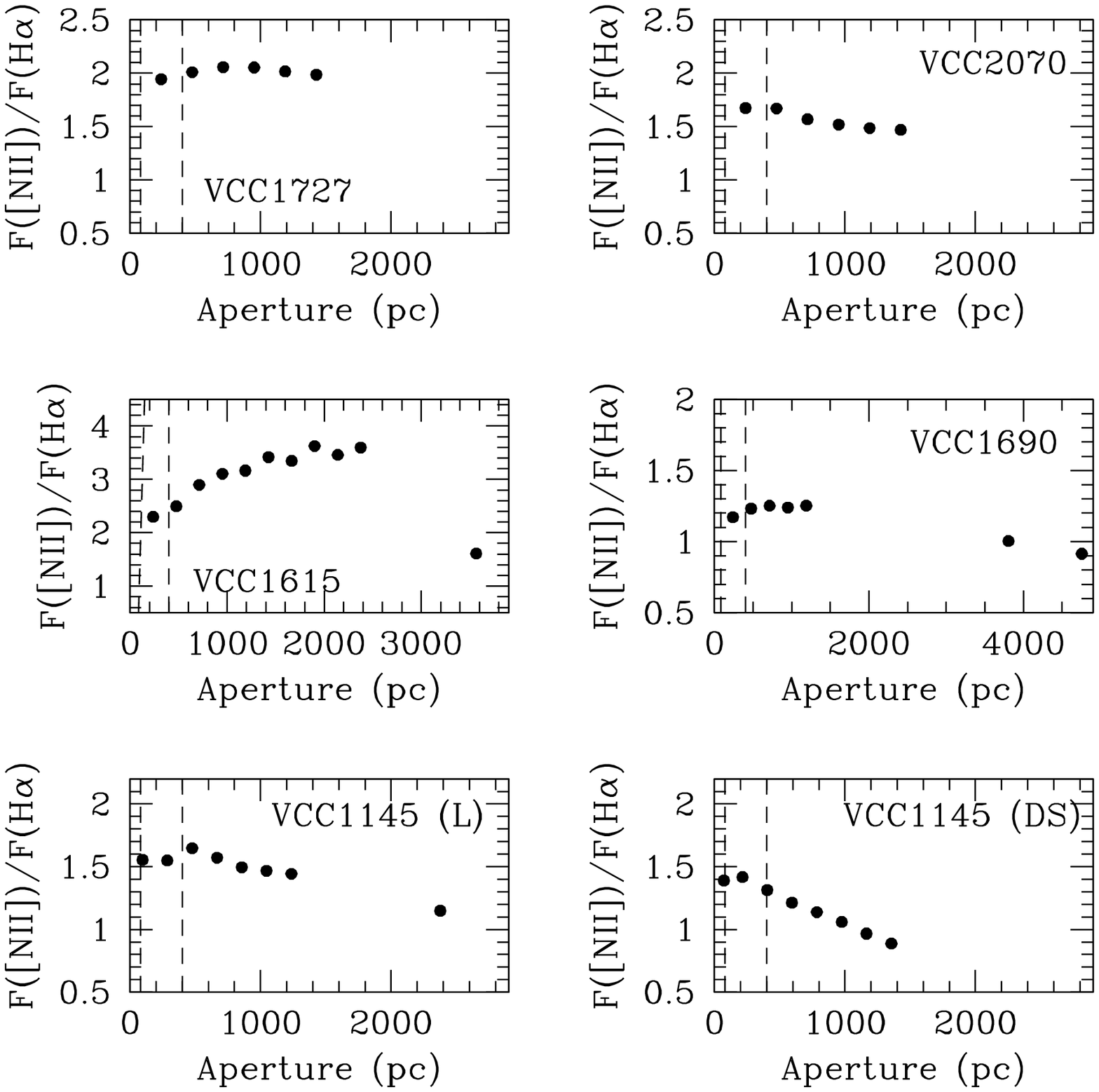}\\
   \caption{The [NII]/H$\alpha$ (H$\alpha$ corrected for absorption)
   ratio in 11 Virgo galaxies with nuclear long-slit spectra taken at Loiano is plotted as a function
   of the aperture in which the spectrum was extracted. The inner vertical dashed
   line corresponds to the seeing. The outer one gives the seeing if the galaxy was at 5 times the Virgo
   distance, i.e. at Coma. One galaxy (e.g. VCC 1145) is given twice: as derived from the Loiano nuclear
   spectrum (L) and from the Drift-scan mode spectrum (DS), the two being consistent.
   }
   \label{apertures}
   \end{center}
   \end{figure*}
\begin{figure*}
\centering
\includegraphics[width=0.7\textwidth]{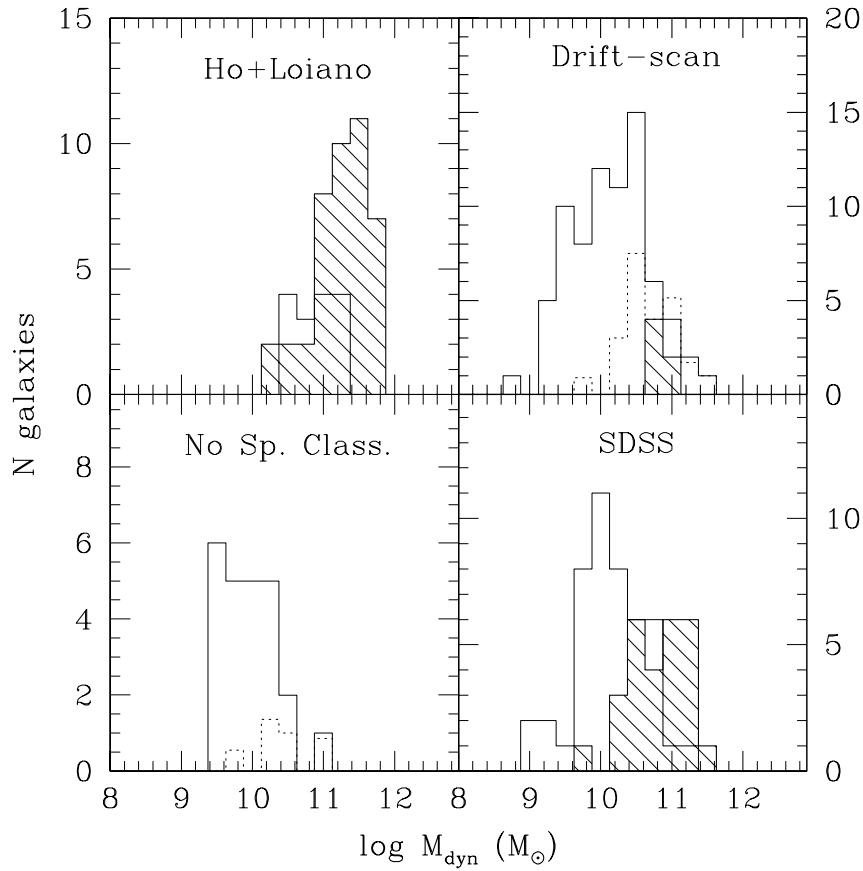}
\caption{\small The observed distribution of AGNs (shaded) and HII
region-like nuclei (thick histogram) as a function of the host
galaxy dynaical mass, given separately for the Ho+Loiano spectra
(top-left); for the SDSS spectra (bottom right); for the Drift-Scan
spectra (top-right). The bottom-left panel shows the luminosity
distribution of 24 galaxies that were not spectroscopically
classified. The dashed histograms show the expected frequency of
AGNs among unclassified and Drift-scan objects, extrapolated from
the distribution of AGNs in the SDSS sample. }\label{corr}
\end{figure*}

\section{Analysis}

Table \ref{tabellone} lists the relevant photometric and spectroscopic parameters derived in this work.
The content of the individual columns is listed at the end of the table.

\subsection{AGN diagnostics}

\begin{table*}
\caption {Spectral classification for all the available data. The 
number of adopted classification separately for each data source is reported 
in parenthesis.}
\label{results}
%\[
%\begin{array}{p{0.5\linewidth}cccccccc}
$\begin{array}{c|cccccc|c}
\hline
\rm Set      & \rm Ho & \rm SDSS & \rm Loiano &\rm mDS & \rm DS & \rm NED & \rm Adopted\\
\hline
\rm Seyfert  &  7~ (7) &   2~ (1) &   0~ (0) &  0~ (0)&    1~ (0) & 12~ (1)&   9 \\ 
\rm Sey/LIN  &  0~ (0) &  15~ (14)&  15~ (5) & 18~ (3)&   26~ (1) &  3~ (0)&  23 \\ 
\rm  LINER   & 17~ (17)&   7~ (6) &   0~ (0) &  0~ (0)&    0~ (0) & 11~ (1)&  24 \\ 
\rm LIN/HII  &  6~ (6) &   8~ (6) &   4~ (3) &  1~ (1)&   20~ (5) &  2~ (0)&  21 \\ 
\rm   HII    & 10~ (10)&  44~ (38)&   8~ (5) &  2~ (0)&  136~ (68)& 13~ (0)& 121 \\ 
\rm   NEL    &  0~ (0) &   8~ (8) &   2~ (0) &  1~ (0)&   10~ (7) &  0~ (0)&  15 \\ 
\hline
\rm  AGN     & 30 &  32 &  19 & 19 &  47& 28 &  77\\
\rm  All     & 40 &  84 &  29 & 22 & 193& 41 & 213\\
\rm  f_{AGN} &0.75& 0.38&0.66 &0.86&0.24&0.68&0.36\\
\hline
\end{array}
$
%\]
\end{table*}

There are various semi-empirical criteria to separate normal from
AGN galaxies and, within the latter, various levels of AGN activity.
They all rely on 2-D line diagnostic diagrams involving
[OIII]${}_{\lambda 5007}$/H$\beta$, [NII]${}_{\lambda 6584}$/H$\alpha$,
[OI]${}_{\lambda 6300}$/H$\alpha$ and ([SII]${}_{\lambda 6717}$+[SII]${}_{\lambda 6731}$)/H$\alpha$
(see e.g. Baldwin, Phillips \& Terlevich, 1981; Veilleux \& Osterbrock 1987;
Kewley et al. 2001; Kauffmann et al. 2003). The reader is referred to
Stasinska et al. (2006)  for a detailed comparison of these diagnostic tools.
In this paper we adopt the following definitions (see Figure \ref{diagno}):

1) for all spectra we adopt [NII]/H$\alpha<0.4$ to unambiguously identify HII-like nuclei,
$0.4<[NII]/H\alpha<0.6$ for transition objects (HII/LIN) and [NII]/H$\alpha>0.6$ for AGNs.

2) if the blue spectrum is available and if [OIII] and H$\beta$ are
detected we split the AGNs among LINERs ([OIII]/H$\beta<3$) and
Seyfert ([OIII]/H$\beta\geq3$), otherwise we classify them as
Sey/LIN.

Table \ref{results} summarizes the results of the adopted
classification. Data sources are listed in order of decreasing
``weight'' from left to right.

\subsection{Spatial distribution of the [NII]/H$\alpha$ ratio}
\label{spatial}

Ionization conditions higher than the normal stellar contribution
characterize the AGNs narrow-line region, with a spatial extent not
exceeding some hundred pc (Bennert et al. 2006). A clear-cut example
is the giant galaxy VCC 1401, hosting an AGN (Ho et al. 1997); our
2-dimensional DS spectrum is shown in Figure \ref{1401} (left
panel). When integrating over the inner 250 pc, the [NII] and
H$\alpha$ lines intensity ratio is typical of AGNs. On the other
hand, the line emission around 2500 pc is typical of normal
star-forming regions (Figure \ref{1401}, right panels). This example
emphasizes that DS spectra may underestimate the number of AGNs in
our sample, when AGN contribution is strongly contaminated from
outer HII-regions.

It was quite surprising however to find that among the 11 AGNs for
which we have long slit nuclear spectra taken at Loiano with
sufficient signal to follow the line intensity at significant
distances from the nucleus (see Figure \ref{apertures}), in only one
galaxy (i.e. VCC 73) the high ionization conditions found in the
nucleus drop to HII-like within 1 kpc. In 90\% of the remaining
objects they prevail up to radii significantly exceeding 1 kpc. Some
galaxies (e.g. VCC 1110) show an even increasing [NII]/H$\alpha$
ratio up to 3 kpc!, in agreement with Veilleux et al. (2003) who
found extended ionization conditions in NGC1365 (see their Fig 6d).
Even in Drift-scan spectra these galaxies would have been recognized
as AGNs, as it is the case for VCC1145, whose spectra are available
in both the nuclear form (L) and Drift-scan mode (DS), showing
consistency.

It is worth thus estimating the number of possibly missed AGNs
because of the use of DS spectra, as illustrated in Figure
\ref{corr}. While the spectra obtained by Ho et al. (1997) and by us
at Loiano were mostly of luminous objects, DS and SDSS spectra cover
all the luminosity range of the galaxies in our sample. If the DS
method would bias against AGNs, a luminosity dependence would be
artificially injected. If we assume that the correct rate of AGNs as
a function of mass is sampled with the nuclear spectra from the
SDSS, we expect that, among all the 193 DS spectra, 87 should have AGNs
signature, that means, the DS data have a 54 \% efficiency in
detecting AGNs. Most of these objects lie in the narrow interval of
mass ($10^{10}-10^{10.5} $ \msol) (see Figure \ref{corr}, top-right
panel). Among the data with no other spectral classification, we
expect 23 AGNs instead of the 6 observed.

In order to reduce this discrepancy, 1-dimensional spectra are
re-extracted from 2-dimensional frames, by integrating the counts on
small apertures ($\sim$ 5 arcsec). These are labeled ``mod-DS''
(modified Drift-Scan) in Table \ref{results} and are obtained in 22
cases out of the 190 available DS spectra. We re-extracted the
spectra only for those galaxies with a significant enhancement in
nuclear continuum emission or with clear inversion of the
[NII]/H$\alpha$ ratio in the two-dimensional spectrum, as in the
case of VCC1401. The spectral classification of VCC524 changed from
HII-like to Transition Objects; VCC596 passed from HII-like to
Sey/LIN AGN; other 5 galaxies changed their classification from
Transition Objects to Sey/LIN subclass. For all the others the
classification remained unchanged. Owing to the existence of more
suitable spectral material (the nuclear spectra, namely from Ho,
SDSS, Loiano) only 81 DS and 4 mod-DS spectra were adopted in the
final classification.

By applying the same argument we expect to find 4 additional AGNs
among the 24 galaxies that remain unclassified because the
spectroscopic material is unavailable.

In conclusion we predict that once nuclear spectra will be available
for the the whole sample, the number of AGNs in the Virgo cluster
would increase at most by $\sim21$ units, over the presently known set of
77.

\section{Results}
\subsection{Nuclear activity versus dynamical mass}

\begin{figure}
   \includegraphics[width=0.5\textwidth]{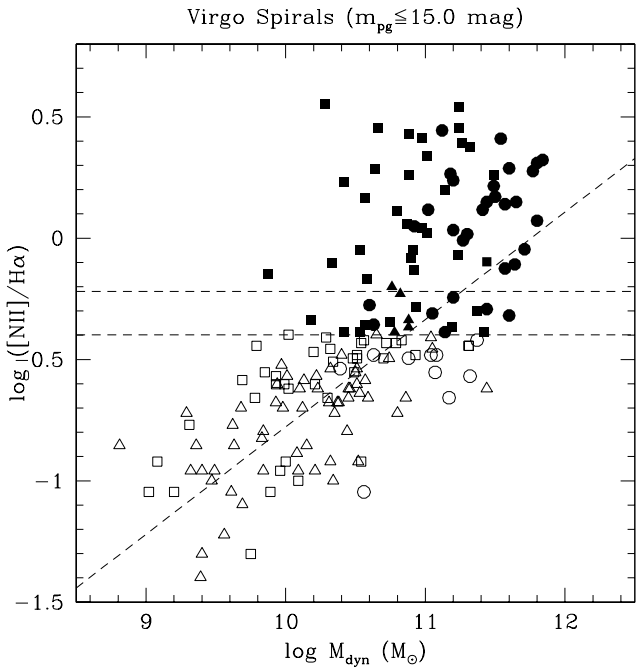}
\caption{\small The dependence of [NII]/H$\alpha$ on the dynamical mass separately for
HII-like nuclei (empty symbols) and AGNs (filled symbols).
}
\label{massdep}
\end{figure}
\begin{figure*}
\begin{center}
\includegraphics[width=0.7\textwidth]{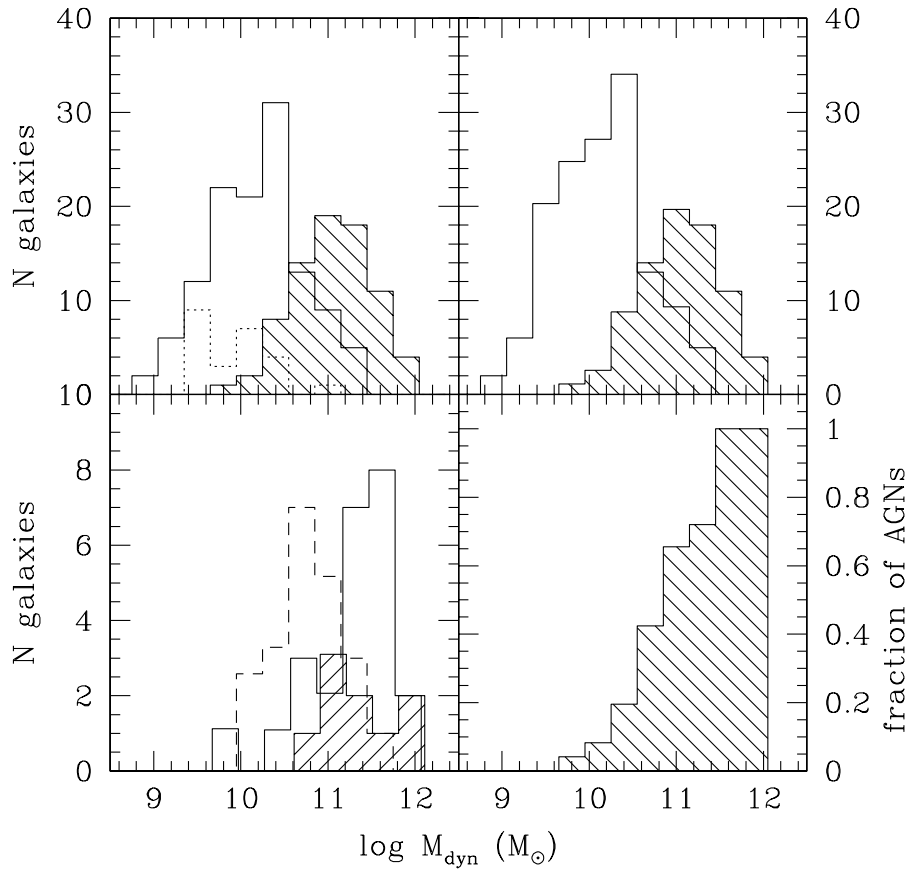}\\
\caption{Histogram of the mass distribution of HII-like (continuous), AGNs (shaded) and
not spectroscopically classified (dotted) (top-left panel);
same  with the ``not classified'' distributed according to the frequency of the classified
(top-right); blow-up of the AGNs divided in Seyfert (shaded), LINERs (continuous),
and Transition (dashed) (bottom-left); fractional differential distribution of AGNs (bottom-right).}
\label{histo}
\end{center}
\end{figure*}
\begin{figure*}
\begin{center}
\includegraphics[width=0.45\textwidth]{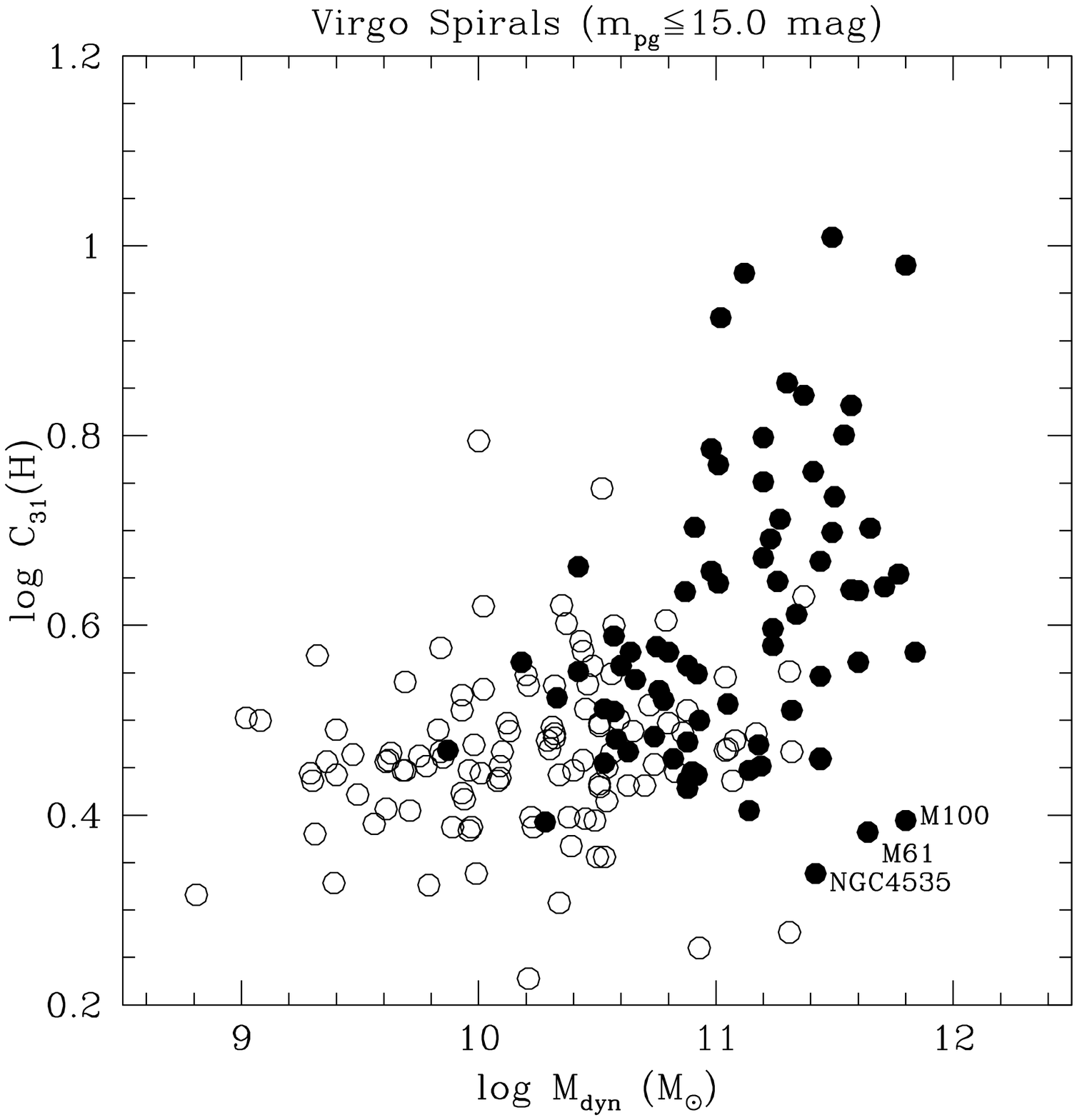}
\includegraphics[width=0.45\textwidth]{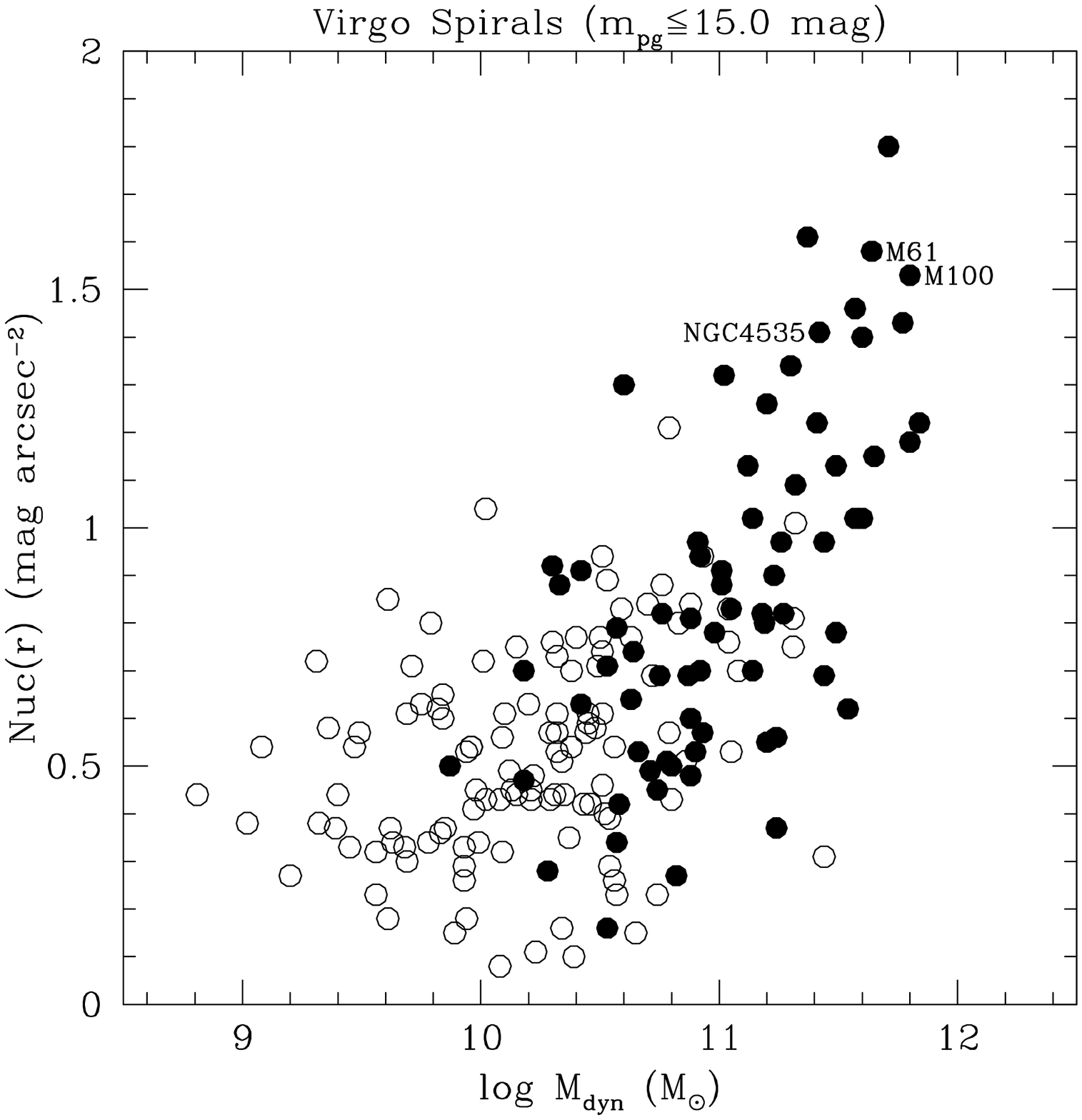}\\
\caption{The Near-IR concentration index (left panel) and the
$r$-band nuclearity index (right panel) as a function of the
dynamical mass. Empty symbols are for HII region-like nuclei, filled
symbols for all kinds of AGN, including transition objects. M61,
M100 and NGC4535 are marked in both diagrams.} \label{mdyn}
\end{center}
\end{figure*}
The data suggest that nuclear activity is very sensitive on the
galaxy dynamical mass ($M_{\rm dyn}$). We observe that the frequency
of AGNs increases very steeply with increasing $M_{\rm dyn}$, a
trend already apparent in Figure \ref{diagno} (the symbol size, set
proportional to $M_{\rm dyn}$, increases toward the right-bottom
angle of the Figure). Table \ref{statistics} clearly shows that the
majority of AGNs (89\%) resides in galaxies more massive than
$10^{10.5}$ \msol.

Figure \ref{massdep} shows the dependence of the [NII]/H$\alpha$
ratio on the mass of the galaxies, that holds already for the HII
like nuclei. This is a  metallicity/luminosity relation well known
since Lequeux et al. (1979); see also Zaritsky (1993) and Gavazzi et
al. (2004). The linear regression for the HII like nuclei is drawn
in Figure \ref{massdep}:
\begin{equation}
\log ([NII]/H_\alpha) = 0.44 \cdot \log M_{\rm dyn}-5.20
\end{equation}
It appears that nuclei classified as AGNs have a [NII]/H$\alpha$
ratio significantly in excess of the relation found for HII nuclei.
This provides a side-aspect, yet important when dealing with imaging
surveys carried on with H$\alpha$ filters (that are broad
enough to contain the [NII] lines braketing H$\alpha$): there is a
substantial, strongly mass dependent contribution from [NII] in the
circumnuclear regions of galaxies. This reminds that significant
(mass dependent) corrections to the H$\alpha$ E.W. are needed to
disentangle the contribution from [NII]; otherwise, it would produce
systematic overestimates of the H$\alpha$ luminosity. This has
relevant consequences for those studies that try to quantify the
amount of circumnuclear star formation as a function of the local
galaxy density using H$\alpha$ imaging surveys (e.g. Moss \&
Whittle, 2005) or studies focused on H$\alpha$ luminosity function
(e.g. Pascual et al. 2001, Fujita et al. 2003, Ly et al. 2006;)

To further study and quantify the frequency of AGNs as a function of
$M_{\rm dyn}$ let us consider the four histograms in Figure
\ref{histo}. The top-left histogram carries the frequency of AGNs,
HII-like+NEL nuclei and the non classified (because their spectra
are unavailable) objects. If we distribute the 24 non-measured ones
according to the frequency of the measured galaxies, we obtain the
top-right histogram showing a clear-cut dichotomy between HII-like
and AGN nuclei at $\log M_{\rm dyn}\sim10.5$.

Once the AGNs are more finely divided into Transition, LINERs and
Seyfert (bottom-left histogram; Sey/LIN objects are not plotted)
there is a barely significant separation of these sub-components as
a function of mass, because of the insufficient statistics (see the
excellent analysis of SDSS data by Kewley et al. 2006). Finally the
bottom-right histogram shows the differential fraction of AGNs
emphasizing that below $10^{10}$ \msol only one late type
galaxy harbors an AGN, while above $10^{11.5}$ \msol all do.

\subsection{Nuclear activity versus galaxy morphology}
\begin{table}
\begin{center}
\caption {AGN, HII and NEL statistics among 213 galaxies with
available spectroscopic classification.} \label{statistics}
%\[
$\begin{array}{cccc}
\hline
\noalign{\smallskip}
                     &\rm AGN~(77) & \rm HII~(121) & \rm NEL~(15)\\
\noalign{\smallskip}
\hline
\log M<10.5          &   9~(12\%)  &  84~(69\%)    &   6~(40\%)   \\
\log M\geq10.5       &  68~(88\%)  &  37~(31\%)    &   9~(60\%)   \\
\hline
\noalign{\smallskip}
\rm Sa-Sb            &   41~(53\%) &   5~(4\%)     &  11~(73\%)   \\
\rm Sbc-Im-BCD       &   36~(47\%) & 116~(96\%)    &   4~(27\%)   \\
\noalign{\smallskip}
\hline
\end{array}$
%\]
\end{center}
\end{table}

As emphasized in Table \ref{statistics}, there is an obvious
tendency for HII region-like nuclei to avoid early-type spirals and
to inhabit late-type spirals (and vice versa for NEL galaxies). On
the other hand, AGNs are almost as frequent in early and late-type
spirals, i.e. the Hubble type does not appear to be the driver of
the difference.

Figure \ref{mdyn} shows the Near-IR $C_{31}$ (left) and the optical ``nuclearity''
(right) plotted as a function of $M_{\rm dyn}$ for galaxies in our sample, divided
in normal (empty) and AGNs (filled).
Albeit a general increase of the $C_{31}$ with $M_{\rm dyn}$,
the two quantities are non-linearly correlated, as pointed out by Boselli et al. (1997),
Gavazzi et al. (2000) and Scodeggio et al. (2002).
There is in fact a number of very large mass ($\log (M_{\rm dyn}/$\msol$)>11.5$)
galaxies with very little bulges ($C_{31}<4$). AGNs inhabit high mass galaxies,
not necessarily high $C_{31}$ (Bulge-to-disk) ones.
Clear-cut examples of bulge-less AGNs are M61, M100 and NGC4535, three Scs with
$\log (M_{\rm dyn}/$\msol$)\sim11.5$ and $C_{31}\sim2.5$.

Conversely the right panel in Figure \ref{mdyn} shows that the
presence of optical nuclei in the host galaxy increases with $M_{\rm
dyn}$ as the fraction of AGNs. An optical nucleus may be due to the
AGN continuum as well as to a steep enhancement in the concentration
of stars. The first one would produce a power-law continuum. No DS
nor SDSS spectrum showed such a behaviour. Spinelli et al. (2006)
published high spatial resolution spectra obtained with HST ($0.2$
arcsec $\approx 16.4$ pc in the Virgo Cloud A) of NGC4450, NGC 4698,
M88 and M90, all having $Nuc(r)>1$. Despite of the high resolution,
no noticeable power law continuum is observed in the
optical band. We thus conclude that the role of the AGN
continuum is commonly negligible in our sample galaxies.
Thus the galaxy dynamical mass seems to be the driver both for the
formation of star nuclei, and for triggering the AGN. Notice that
the massive, bulge-less Sc M61, M100 and NGC4535 have significant
nuclei ($Nuc(r)\sim1.50$).

\begin{figure}
   \includegraphics[width=0.5\textwidth]{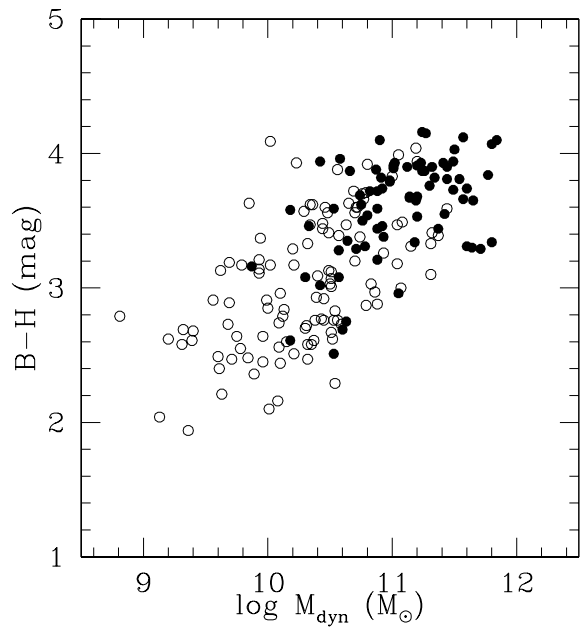}
\caption{\small The dependence of B-H color on the dynamical mass of
late-type galaxies with HII-like nuclei (empty symbols) and AGNs
(filled symbols). } \label{colmag}
\end{figure}

Obric et al. (2006) suggest that AGN host galaxies are mainly redder
than star forming galaxies. This is interpreted as an evidence of
feedback by the AGN on the host galaxy. AGNs do reside in red
galaxies, as shown in Figure \ref{colmag}, but the existing and
well-known color-magnitude correlation stands for both active and
non-active galaxies. The role of the AGN in halting star formation
in the host galaxy is still debated.

\subsection{Environmental dependencies}

Kauffmann et al. (2004) find that in the SDSS the fraction of high
luminosity AGNs (those with L[OIII]$>10^7 \rm L_\odot$), predominant
among type 2 Seyfert galaxies, depends on the environment, i.e. it
decreases with increasing galaxy density, mimiking the behaviour of
emission line galaxies. Conversely these authors find, in agreement
with Miller et al. (2003), that the frequency of AGNs in general
(dominated in number by low-luminosity LINERs) does not depend on
the environment.

With a sample like the presently analyzed one, that is entirely made
of galaxies members of a rich cluster, it is impossible to ensure
that the estimate of the frequency of AGNs is not biased by the
density of this particular environment. Neither we can perform a
detailed study of environmental effects for each AGN subclass
separately, due to the poor statistics.

Nevertheless we notice firstly that the overall fraction of AGNs
associated with massive galaxies in the SDSS ($> 80$ \% for $\log
(M_{\rm stars}/$\msol$)>11$, Kauffmann et al. 2003), is consistent
with the one found in Virgo (see Figure \ref{histo})\footnote{We
transform $M_{\rm stars}$ used by the SDSS into $M_{\rm dyn}$
(within the optical radius) adopted in the present paper, using
$\log (M_{\rm stars}/$\msol$)=1.41 \times \log (M_{\rm
dyn}/$\msol$)-5.1$. This transformation is the linear correlation
between $\log M_{\rm dyn}$ and $\log M_{\rm stars}$ of Bruzual \&
Charlot synthesis population models fitted to the SEDs of the
sampled spirals}. Such a result is in contrast with Popesso \&
Biviano (2006). If we adopt an average galaxy velocity dispersion
around 600 km/s for each Virgo Cloud (see Gavazzi et al. 1999), cut
our sample at $M_{\rm dyn} \approx 6.3 \cdot 10^{10}$ \msol (roughly
corresponding to $M_{\rm pg} \approx -20$), and adopt the same
limits in classifing the nuclear activity, our AGN fraction reaches
60\%, nearly a 2 $\sigma$ deviation from their correlation. We argue
that the reason of such a difference resides in the sample
selection. Unlike Kauffmann et al. (2003), the sample of Popesso \&
Biviano has no constraints on line fluxes. Thus, it includes lots of
No Emission Line galaxies, especially among early-type, gas
poor galaxies for which the BPT classification may be unavailable.

\begin{figure}
\centering
\includegraphics[width=0.5\textwidth]{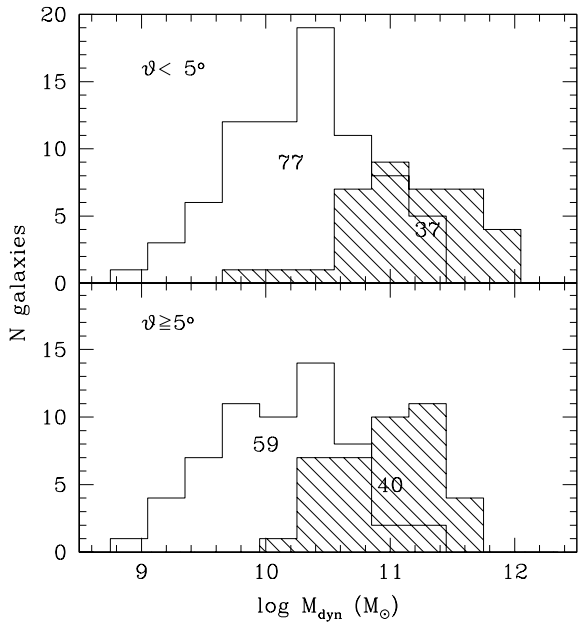}
\caption{\small The observed distribution of AGNs (shaded) and HII
region-like+NEL nuclei (thick histogram) as a function of $M_{\rm
dyn}$, given separately for galaxies found within (top) or outside
(bottom) 5 degrees projected radial distance from M87. }\label{def}
\end{figure}

\begin{figure}
\centering
\includegraphics[width=0.5\textwidth]{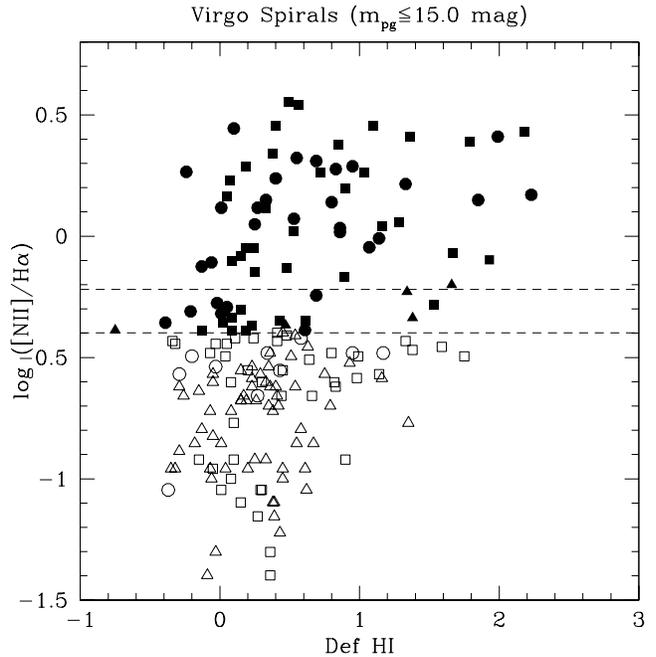}
\caption{\small [NII]/H$\alpha$ ratio plotted versus Hydrogen
deficiency, as defined in Haynes \& Giovanelli (1984). The AGN
fraction increases only slightly from non-deficient to deficient
galaxies.}\label{defhi}
\end{figure}

Secondly we remark that, if we divide the Virgo sample in two
subsamples composed of galaxies in and outside 5 degrees ($\sim 1.5$
Mpc) projected radial distance from M87, we obtain that the
percentage of AGNs varies from $32\pm 8$\% ``in" to $40\pm 10$\%
``out", i.e. not significantly. The lack of a environmental
dependence of our sample AGNs is illustrated in Figure \ref{def}.
Similarly by dividing the sample using an independent probe of the
environmental influence, as represented by the Hydrogen deficiency
parameter (see the parameter definition in Haynes \& Giovanelli,
1984 and Boselli \& Gavazzi, 2006), we obtain that the percentage of
AGNs varies from $41\pm 13$\% among highly deficient objects
($Def_{HI}>0.5$), to $27\pm 7$\% among the unperturbed objects
($Def_{HI}\leq0.5$), i.e. barely significantly (see figure
\ref{defhi}). We thus conclude that the frequency of AGNs is not
strongly environment dependent.

\subsection{Indirect estimate of black hole masses and Eddington ratios}

\begin{figure}
   \includegraphics[width=0.5\textwidth]{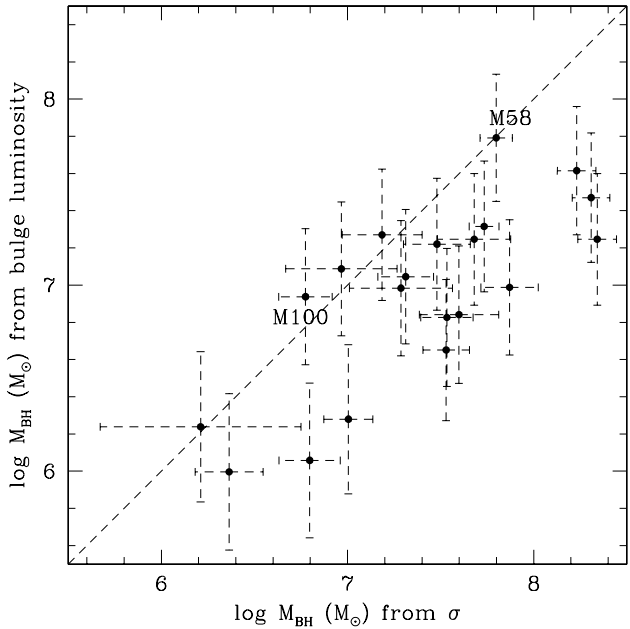}
\caption{\small Comparison between $M_{\rm BH}$ estimated using the
$M_{\rm BH}$ -- $\sigma$ relation (Tremaine et al., 2002) ($x$ axis)
and $M_{\rm BH}$ from the $M_{\rm BH}$ -- $L_{\rm bulge}$ (Marconi
\& Hunt, 2003) ($y$ axis). Uncertainties in the relations are
accounted for in our error bars; for $L_{\rm bulge}$ we assumed a 30
\% error both on the Bulge to Total value and on total NIR
Luminosity.} \label{mbh}
\end{figure}

With the current data it is not possible to directly infer the mass
of the black holes that power the AGNs observed in our sample. We
may nevertheless consider the possibility of extending the known
scale relations valid for spheroids to the bulges and optical nuclei
of our late-type galaxies. The relations used in this section are
the $M_{\rm BH}$ -- $\sigma$ (Tremaine et al., 2002 and references
therein) and the $M_{\rm BH}$ -- $L{}_{\rm bulge}$(H) (Marconi \&
Hunt, 2003). Data on the stellar velocity dispersion $\sigma$ are
taken from hypercat\footnote{http://leda.univ-lyon1.fr} when
available for the AGN sample. The values for $\sigma$ are estimated
using various techniques (mainly the Fourier quotient method). Data
on the bulge luminosity are taken from Gavazzi et al. (2000). The
bulge luminosity is estimated from the deconvolution of the galaxy
light profile in H-band. The light profile is modeled as a
de Vaucouleurs or an exponential curve for the bulge and an
exponential law for the disk component. Since the AGN continuum flux
is found to be negligible in our sample galaxies, and the fit
procedure is performed only out of the first two seeing radii, we
disregarded the AGN contribution in the fit. Typical values
for the Bulge to Total luminosity ratio (B/T) varies between few \%
to 30\%. A reasonable estimate of the bulge luminosity error is
around 30\%. Twenty and 64 out of 77 AGNs in our sample have data
for these two estimates to be performed; this subsample includes,
for example, M100 with its optical nucleus and M58 which hosts a
prominent bulge. Figure \ref{mbh} shows the comparison between the
black hole mass estimates obtained with the two methods. The slight
($0.4$ dex) offset between the two estimates may depend on the
heterogeneous data sources adopted for $\sigma$ estimates, or on the
bulge luminosity underestimate due to inclination. Since we want to
study the black-hole masses and the Eddington ratios of Virgo AGNs
only from a statistical point of view, and the data dispersion is
higher, such a offset is irrelevant for the purposes of our
analysis.
We notice that the two methods provide consistent values of the
black hole mass for M100 as well as for M58. This result is
surprising given the difference between M100 and M58, the first
hosting a pseudobulge/optical nucleus (Kormendy \& Kennicutt, 2004;
see also Allard et al., 2006) while the second a massive extended
bulge: this coincidence highlights the close underlying connection
between the stellar velocity dispersion and the nuclear or bulge
luminosity in these two galaxies.

As shown in Figure \ref{mbh} the black hole masses never exceed a
few $10^8$ solar masses, and a number of galaxies hosts relatively
light black holes ($M_{\rm BH} \lsim 10^7 $ \msol). These black
holes appear to fill the low mass end of the scaling relation purely
sampled in our Local Universe.

\begin{figure}
   \includegraphics[width=0.5\textwidth]{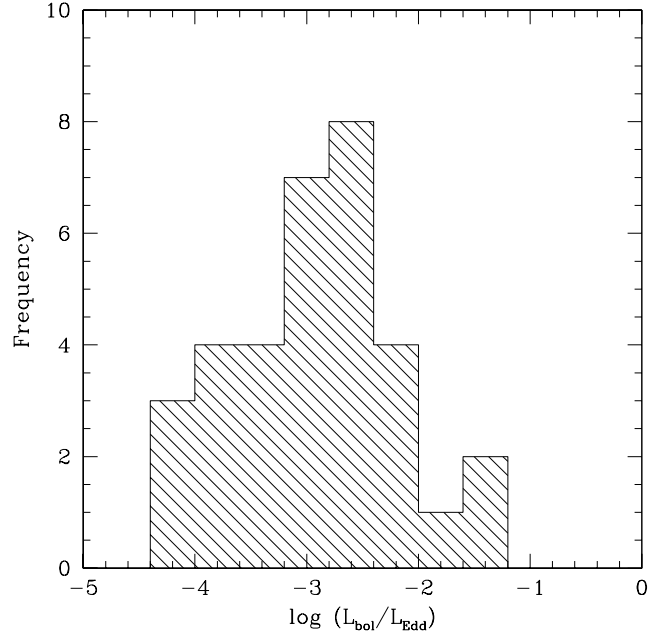}
\caption{\small Eddington ratio distribution in a subset of AGNs in
our sample. The Marconi \& Hunt relation was applied to estimate
black hole masses and Eddington luminosities.} \label{accret}
\end{figure}

Black hole masses may provide some hints on the Eddington ratios of
the AGNs if we have a way to estimate the bolometric luminosity of
the active BH. Since most of our AGN are LINERs, it is difficult to
disentangle the emission of the AGN from that of the stellar
continuum in the optical band. As proposed in Heckman et al. (2004)
(see also Kauffmann \& Heckman, 2005), we can estimate the AGN
bolometric luminosity from the [OIII] emission line flux. We want to
stress that this is a rough estimate of bolometric luminosities,
since a good spectrometric calibration is often hard to achieve,
because of the diffuse nature of the source, and contamination from
other ionizing radiation sources is often present. Combining the
data on the [OIII] flux from Ho and SDSS, and matching them with the
subsample of the AGN BH masses, we infer the bolometric to Eddington
luminosity ratios of 33 of our AGNs (we could not use Loiano spectra
since they do not include [OIII] line, neither drift scan spectra
since they lack of an absolute spectrophotometric calibration). The
mean error of Heckman's bolometric correction is $0.38$ dex, thus
our Eddington ratios mean error is nearly $0.5$ dex. These ratios
are found to fall in the range $10^{-5}-10^{-2}$ with a mean around
$0.001$, indicating that these AGN are accreting at very low rates
(see Figure \ref{accret}). The only exceptions are NGC4388, with
Eddington ratio of 0.06 which reflects the occurrence of ejection in
this source (see Yoshida et al., 2004), and NGC4123, a LIN/HII
object, in which O,B-stars ionization is probably significant.

\section{Conclusions}

The present study focused on the nuclear activity of spiral galaxies
in the Virgo cluster, as representative of our Local Universe. This
paper provides the first census of AGNs associated to the cluster's
spirals, complete down to the $15^{\rm th}$ magnitude. This allowed
the identification of a number of low luminosity AGNs even in late
type spirals that show a bright optical stellar nucleus rather than
a prominent bulge: M100 hosts a LINER and offers a clear example.

A first result is that the AGNs, and among them LINERs, are
primarily associated to the spirals with dynamical mass in excess of
$M\gsim10^{10} $ \msol. This result is in close agreement with
Kauffmann et al. (2003) who found that AGNs in the SDSS are hosted
in galaxies (with mean redshift $\sim 0.1$) as massive as ours.

In particular, we found the fraction of AGNs in late-type galaxies
more massive than $\sim10^{8.5}$ \msol is $77/237 \approx 32$  \%.
This fraction steeply increases with the galaxy mass: if we consider
only galaxies with $M_{\rm dyn}>10^{11.5}$ \msol, the AGN fraction
approaches 100 \%.

A second result is that the presence of nuclear activity is
insensitive to the morphological type of the host spiral, as it is
equally present in Sa - Sb and later types. Thus, M100 is not an
isolated case. AGNs are often associated to prominent stellar
nuclei. In addition, the nuclear activity displays little or no
environmental density dependence inside the cluster, neither
as a function of the host galaxy position within the cluster, nor of
the neutral gas deficiency. Gas accretion onto central BHs seems independent
of the gaseous content.

As already pointed out, the nuclear activity  occurs in spiral
galaxies with dynamical masses above  $\sim 10^{10} $ \msol, and the
fraction of $M_{\rm dyn}$ in the spheroidal component in spirals
varies between a few \% to 30 \%. If we apply the known black hole
scaling relations, obeyed in ellipticals, to the bulges and optical
nuclei of our sample (thus assuming self-similarity among early and
late type galaxies), this translates into a lower limit for the
black hole mass of $\sim 10^6 $ \msol. Based on AGN luminosity
function evolution, Marconi et al. (2006) have suggested that also
the black holes grow following the galaxy downsizing, i.e. massive
black holes grow faster and earlier with cosmic time than the
lighter ones. If black holes lighter than $10^6 $ \msol are hosted
in spirals with less prominent nuclei/bulges, we would have expected
them still in their active phase in the galaxies inside our census.
The lack of AGNs in lower mass spirals may be a hint that such
lighter BHs did not find yet the right conditions to grow or the
environment to form. This dichotomy on the BH behaviour around $10^6
$ \msol has to be considered valid at a statistical level.

It is interesting to notice that ACS imaging surveys on elliptical
galaxies have shown that a sizeable fraction of low and intermediate
luminosity ellipticals contain stellar cusps at their center with
masses in the range expected from the lower-end extrapolation of the
$M_{\rm BH}$ -- Magnitude correlation (Wehner \& Harris, 2006;
Ferrarese et al. 2006b). Above $10^{10}$ \msol ellipticals host
exclusively massive black holes. Similarly, in our sample of
spirals, we have not found at statistical level an active AGN below
such a characteristic dynamical mass once we correct for the total
to bulge ratio. This is a hint that below a critical mass for the
hosting spheroidal component (either a bulge or a stellar nucleus),
lighter black holes become rarer.

Only a relatively small sample of black holes in spiral galaxies
have a direct mass measure. In this paper we found the
occurrence of BHs at the center of spiral galaxies
disregarding of the galaxy morphology, and we provided an
estimate of the black hole mass under the assumption of
self-similarity between the bulges and stellar nuclei of our spirals
with the sample of quiescent black hole hosted in ellipticals. We do
not know if this similarity applies, since either in the
black hole mass as well as in the mass accretion rate there might be
a dependence on the angular momentum of the host spiral that can
affect black hole grow. Further high resolution observations focused
on black hole mass measures in late-type galaxies may provide clues
on the applicability of the $M_{\rm BH}$ -- bulge relations and the
role of the both the Bulge to Total ratio and the dynamic of the gas
and stars in the host galaxy.

\subsection*{Acknowledgements}
We thank  F. Haardt and  A. Treves for useful discussions. G.G.
thanks A. A. for inspiring hints. This research has made extensive
use of the GOLDMine Database (Gavazzi et al. 2003). We acknowledge
the usage of the HyperLeda database (Paturel et al. 2003). This
research has made use of the NASA/IPAC Extragalactic Database (NED)
which is operated by the Jet Propulsion Laboratory, California
Institute of Technology, under contract with the National
Aeronautics and Space Administration.

%\clearpage
\onecolumn
\setlength{\LTcapwidth}{16truecm}

\begin{tiny}
\begin{longtable}{cccccccccccc}
\caption {spectroscopic results}\label{tabellone}\\
\hline
\noalign{\smallskip}
VCC/CGCG & NGC & TYPE   &$m_{\rm pg}$ &$D$& log$L_H$& $C_{31}$(H)   &  $Nuc(r)$          & [NII]/H$\alpha$ &  [OIII]/H$\beta$ &  Spec Class  &   Ref  \\
         &     &        &     mag     &Mpc& $L\odot$&               &  mag arcsec$^{-2}$ &                 &                  &              &        \\
    (1)  & (2) &  (3)   &     (4)     &(5)&  (6)    &    (7)        &    (8)             &        (9)      &     (10)         &   (11)       &   (12) \\
\noalign{\smallskip}
\hline
\noalign{\smallskip}
\endhead
\noalign{\smallskip}
\hline
\endfoot
1     &     -  & BCD   &  14.78 & 32 &  9.27 &  3.36   &  0.29   &  0.27   &  0.48     &     HII   &   SDSS \\
10    &     -  & BCD   &  14.75 & 32 &  9.27 &  2.65   &  0.33   &  0.25   &  0.71     &     HII   &   SDSS \\
15    &     -  & Sa    &  14.70 & 32 &   -   &   -     &   -     &  0.45   &   -       &   LIN/HII &   SDSS \\
24    &     -  & BCD   &  14.95 & 32 &  9.34 &  6.23   &   -     &  0.12   &  2.37     &     HII   &   SDSS \\
25    &  4152  &  Sc   &  12.46 & 32 & 10.39 &  3.29   &  0.83   &  0.32*  &  0.51*    &   LIN/HII &   Ho   \\
47    &  4165  & Sa    &  14.20 & 32 & 10.09 &  3.78   &  0.69   &  0.45   &	-      &   LIN/HII &   SDSS \\
58    &     -  & Sb    &  13.17 & 32 & 10.17 &  2.79   &   0.8   &  0.38   &	-      &     HII   &   SDSS \\
66    &  4178  &  Sc   &  11.89 & 17 & 10.22 &  3.24   &  0.84   & 0.19*   &   1.37*   &     HII   &   Ho   \\
67    &     -  &  Sc   &  13.98 & 32 &  9.86 &  5.55   &   0.4   & 0.12*   &   -       &     HII   &   DS   \\
73    &  4180  & Sb    &  13.35 & 32 & 10.58 &  3.79   &  0.56   &  2.84   &   -       &   Sey/LIN &   LOI  \\
87    &     -  & Sm    &  15.00 & 17 &  8.36 &  3.18   &  0.38   &  0.09   &  2.09     &     HII   &   SDSS \\
89    &  4189  &  Sc   &  12.53 & 32 & 10.65 &  1.89   &  0.81   &  0.36   &  0.14     &     HII   &   SDSS \\
92    &  4192  & Sb    &  10.92 & 17 & 10.99 &  5.04   &  1.15   &  1.29   &   -       &     LIN   &   Ho   \\
97    &  4193  &  Sc   &  13.20 & 32 & 10.53 &  2.83   &   0.8   &  0.43   &  0.42     &  LIN/HII  &   SDSS \\
105   &     -  &  Sd   &  13.68 & 32 &   -   &   -     &  0.64   &   0.27* &   -       &     HII   &   DS   \\
119   &     -  &  Sc   &  14.76 & 32 &   -   &   -     &  0.19   &  0.08   &  1.28     &     HII   &   SDSS \\
120   &  4197  &  Scd  &  13.47 & 32 & 10.20 &  3.07   &  0.51   &  0.22*  &  1.14*    &     HII   &   DS   \\
126   &     -  &  Sd   &  14.42 & 17 &   -   &   -     &   -     &    -    &   -       &     NEL   &   SDSS \\
131   &     -  &  Sc   &  14.34 & 17 &  9.52 &  3.64   &  0.47   &  0.46   &   -       &  LIN/HII  &   SDSS \\
135   &     -  & S/BCD &  14.81 & 32 &  9.82 &  3.61   &  0.58   &  0.32   &   -       &     HII   &   SDSS \\
145   &  4206  &  Sc   &  12.77 & 17 &  9.98 &  3.73   &  0.74   &  1.93   &   -       &     LIN   &   SDSS \\
152   &  4207  &  Scd  &  13.48 & 17 &  9.90 &  3.54   &  0.54   &  0.38   &  0.45     &     HII   &   SDSS \\
157   &  4212  &  Sc   &  11.50 & 17 & 10.48 &  2.54   &   0.7   &  0.35*  &   -       &   LIN/HII &   Ho   \\
162   &     -  &  Sd   &  14.41 & 17 &  9.23 &  2.44   &  0.15   &  0.09   &  1.83     &     HII   &   SDSS \\
167   &  4216  & Sb    &  10.97 & 17 & 11.14 &  9.54   &  1.18   &  3.27   &   -       &     LIN   &   Ho   \\
170   &     -  &  Sd   &  14.56 & 17 &  9.03 &  2.80   &  0.61   &  0.26   &  0.49     &     HII   &   SDSS \\
172   &     -  & BCD   &  14.50 & 32 &  9.30 &  2.80   &  0.54   &  0.11   &  4.28     &     HII   &   SDSS \\
187   &  4222  &  Scd  &  13.91 & 17 &  9.57 &  2.44   &  0.11   &  0.24*  &  0.88*    &     HII   &   DS   \\
199   &  4224  & Sa    &  12.95 & 32 & 10.83 &  4.99   &  0.78   &    -    &   -       &   Sey/LIN &   LOI  \\
213   &     -  & S/BCD &  14.26 & 17 &  9.21 &  2.94   &   0.5   &  0.71   &   -       &     LIN   &   SDSS \\
221   &  4234  &  Sc   &  13.43 & 32 &  9.88 &  2.60   &  0.39   &  0.37   &  0.48     &     HII   &   SDSS \\
222   &  4235  & Sa    &  12.62 & 32 & 10.88 &  6.32   &  0.62   &    -    &   -       &     Sey   &   Ho   \\
224   &     -  &  Scd  &  14.70 & 17 &  9.19 &  2.89   &  0.37   &  0.28   &  0.62     &     HII   &   SDSS \\
226   &  4237  &  Sc   &  12.53 & 17 & 10.26 &  2.77   &   0.7   &  0.74   &   -       &   Sey/LIN &   SDSS \\
227   &     -  & Sdm   &  14.90 & 32 &  9.44 &  3.17   &   -     &    -    &   -       &     NEL   &   SDSS \\
234   &  4241  & Sa    &  12.99 & 32 & 10.68 &  4.09   &   -     &  7.69   &   -       &   Sey/LIN &   MDS  \\
241   &     -  &  Sd   &  14.60 & 17 &   -   &   -     &  0.35   &  0.04   &  3.37     &     HII   &   SDSS \\
267   &     -  &  Sbc  &  13.82 & 23 &  9.72 &  2.50   &  0.54   &  0.21*  &   -       &     HII   &   DS   \\
289   &  4252  &  Sc   &  14.81 & 32 &  9.33 &  2.18   &  0.34   &  0.25   &  3.85     &     HII   &   SDSS \\
307   &  4254  &  Sc   &  10.43 & 17 & 10.94 &  3.64   &  1.02   &  0.33*  &  0.22*    &   LIN/HII &   Ho   \\
315   &     -  & Sa    &  14.98 & 32 &  9.70 &   -     &   -     &    -    &   -       &     NEL   &   SDSS \\
318   &     -  &  Scd  &  14.01 & 32 &  9.18 &  2.93   &   0.6   &  0.16*  &  1.68*    &     HII   &   DS   \\
323   &  4257  & Sa    &  14.91 & 32 & 10.00 &  3.49   &  0.53   &  2.84   &   -       &     LIN   &   SDSS \\
324   &     -  & BCD   &  14.78 & 17 &  9.09 &  2.90   &  0.63   &  0.05   &  3.88     &     HII   &   SDSS \\
340   &     -  & BCD   &  14.43 & 32 &  9.43 &  2.83   &  0.32   &  0.10   &  2.37     &     HII   &   SDSS \\
341   &  4260  & Sa    &  12.70 & 23 & 10.60 &  4.43   &  0.97   &  2.46   &   -       &   Sey/LIN &   SDSS \\
358   &  4264  & Sa    &  13.80 & 23 & 10.03 &  3.76   &   -     &    -    &   -       &     NEL   &   SDSS \\
382   &  4273  &  Sc   &  12.37 & 32 & 10.65 &  3.56   &  0.75   &  0.36   &   -       &     HII   &   LOI  \\
386   &  4277  & Sa    &  14.47 & 32 & 10.05 &  3.46   &   -     &    -    &   -       &     NEL   &   DS   \\
393   &  4276  &  Sc   &  13.25 & 23 &  9.85 &  2.69   &  0.74   &  0.25*  &   -       &     HII   &   DS   \\
404   &     -  &  Scd  &  15.00 & 17 &  9.36 &  3.41   &  0.43   &  0.40   &   -       &     HII   &   SDSS \\
415   &     -  &  Sd   &  14.82 & 23 &   -   &   -     &  0.28   &    -    &   -       &     HII   &   DS   \\
459   &     -  & BCD   &  14.95 & 17 &  8.74 &  3.09   &  0.44   &  0.11*  &   2.88*   &     HII   &   DS   \\
460   &  4293  & Sa    &  11.20 & 17 & 10.78 &  3.52   &  0.97   &  4.19   &   -       &     LIN   &   Ho   \\
465   &  4294  &  Sc   &  12.62 & 17 &  9.88 &  2.82   &  0.29   &  0.12   &  1.94     &     HII   &   SDSS \\
483   &  4298  &  Sc   &  12.08 & 17 & 10.42 &  3.01   &   0.7   &  0.35   &  0.23     &     HII   &   Ho   \\
491   &  4299  &  Scd  &  12.86 & 17 &  9.42 &  2.73   &  0.08   &  0.13*  &  2.05*    &     HII   &   DS   \\
492   &  4300  & Sa    &  13.76 & 23 & 10.21 &  4.32   &  0.69   &  1.14   &   -       &   Sey/LIN &   LOI  \\
497   &  4302  &  Sc   &  12.55 & 17 & 10.58 &  3.95   &  0.37   &  3.49   &   -       &   Sey/LIN &   SDSS \\
508   &  4303  &  Sc   &  10.17 & 17 & 10.98 &  2.41   &  1.58   &  0.41*  &  0.28*    &     LIN   &   Ho   \\
509   &     -  & Sdm   &  14.98 & 23 &   -   &   -     &  0.54   &    -    &   -       &     HII   &   DS   \\
522   &  4305  & Sa    &  13.19 & 17 &  9.83 &  3.00   &   -     &    -    &   -       &     NEL   &   SDSS \\
524   &  4307  &  Sbc  &  12.79 & 23 & 10.27 &  3.16   &  0.57   &  0.52   &  0.28     &   LIN/HII &   MDS  \\
534   &  4309  & Sa    &  13.59 & 23 & 10.10 &  3.63   &  0.88   &    -    &   -       &     NEL   &   DS   \\
552   &     -  &  Sc   &  13.61 & 17 &  9.05 &  2.54   &  0.71   &    -    &   -       &     HII   &   DS   \\
559   &  4312  & Sab   &  12.56 & 17 & 10.22 &  3.61   &  0.48   &  0.46*  &   -       &   LIN/HII &   DS   \\
568   &     -  & S..   &  14.91 & 23 &   -   &   -     &  0.28   &  0.11*  &   -       &    HII    &   DS   \\
570   &  4313  & Sab   &  12.73 & 17 & 10.32 &  4.54   &   -     &  2.26   &   -       &   Sey/LIN &   SDSS \\
576   &  4316  &  Sbc  &  13.70 & 23 & 10.24 &  2.79   &  0.53   &  0.83   &   -       &   Sey/LIN &   SDSS \\
596   &  4321  &  Sc   &  10.11 & 17 & 11.14 &  2.48   &  1.53   &  0.66   &   -       &     LIN   &   Ho   \\
613   &  4324  & Sa    &  12.60 & 17 & 10.35 &  4.41   &  0.91   &  2.34   &   -       &   Sey/LIN &   LOI  \\
630   &  4330  &  Sd   &  13.10 & 17 &  9.91 &  3.98   &  0.23   &  0.26*  &   -       &     HII   &   DS   \\
655   &  4344  & S/BCD &  13.21 & 17 &  9.56 &  2.50   &  0.48   &  0.27*  &  0.17*    &     HII   &   DS   \\
656   &  4343  & Sb    &  13.14 & 23 & 10.35 &  5.88   &  0.88   &  1.05   &   -       &   Sey/LIN &   LOI  \\
664   &     -  &  Sc   &  13.50 & 17 &  8.95 &  2.55   &  0.85   &  0.09*  &  2.68*    &     HII   &   DS   \\
667   &     -  &  Sc   &  14.24 & 23 &  9.78 &  3.74   &  0.57   &  0.16*  &   -       &     HII   &   DS   \\
675   &     -  & Sa    &  15.00 & 17 &  8.54 &   -     &  0.27   &  0.09   &  2.77     &     HII   &   SDSS \\
688   &  4353  &  Sc   &  13.94 & 23 &  9.68 &  2.03   &  0.16   &  0.10*  &   -       &     HII   &   DS   \\
692   &  4351  &  Sc   &  12.93 & 17 &  9.64 &  2.95   &  0.76   &  0.22   &  0.28     &     HII   &   SDSS \\
699   &     -  & Pec   &  14.22 & 23 &  9.71 &  4.00   &  0.35   &  0.21*  &  1.31*    &     HII   &   DS   \\
713   &  4356  &  Sc   &  14.04 & 23 & 10.16 &  2.88   &  0.27   &  0.59*  &   -       &  LIN/HII  &   DS   \\
737   &     -  & S/BCD &  14.94 & 17 &  8.73 &  2.13   &  0.37   &    -    &   -       &     HII   &   DS   \\
739   &     -  &  Sd   &  14.37 & 17 &  8.65 &  2.40   &  0.72   &  0.17   &  1.11     &     HII   &   SDSS \\
785   &  4378  & Sa    &  12.16 & 17 & 10.46 &  9.36   &  1.13   &    -    &   -       &     Sey   &   Ho   \\
787   &  4376  &  Scd  &  13.69 & 23 &  9.65 &  3.11   &  0.44   &  0.21*  &   -       &     HII   &   DS   \\
792   &  4380  & Sab   &  12.36 & 23 & 10.66 &  3.24   &  1.09   &  2.38   &   -       &   Sey/LIN &   SDSS \\
809   &     -  &  Sc   &  14.55 & 17 &  9.27 &  3.24   &  0.26   &  0.21*  &  0.85*    &     HII   &   DS   \\
827   &     -  &  Sc   &  13.76 & 23 & 10.14 &  3.14   &  0.43   &  0.19*  &  1.80*    &     HII   &   DS   \\
836   &  4388  & Sab   &  11.83 & 17 & 10.54 &  4.69   &  0.55   &  0.56   &  11.34    &     Sey   &   Ho   \\
848   &     -  & 16    &  14.72 & 23 &  8.70 &  2.86   &  0.58   &  0.14*  &  2.05*    &     HII   &   DS   \\
849   &  4390  &  Sbc  &  13.27 & 23 &  9.79 &  2.49   &  0.59   &  0.22*  &  0.94*    &     HII   &   DS   \\
851   &     -  &  Sc   &  14.14 & 23 &  9.80 &  3.45   &  0.42   &  0.24*  &  0.77*    &     HII   &   DS   \\
857   &  4394  & Sb    &  11.76 & 17 & 10.54 &  5.64   &  1.26   &  1.02   &   -       &     LIN   &   Ho   \\
859   &     -  &  Sc   &  14.61 & 17 &  9.62 &  2.47   &  0.28   &  3.58   &   -       &   Sey/LIN &   SDSS \\
865   &  4396  &  Sc   &  13.02 & 17 &  9.69 &  4.18   &  0.44   &  0.19*  & 0.97*     &     HII   &   DS   \\
873   &  4402  &  Sc   &  12.56 & 17 & 10.39 &  2.95   &  0.53   &  0.35*  &   -       &     HII   &   DS   \\
874   &  4405  &  Sc   &  12.99 & 17 &  9.97 &  2.70   &  0.77   &  0.37   &   -       &     HII   &   Ho   \\
905   &     -  &  Sc   &  13.42 & 23 &  9.66 &  3.06   &  0.53   &  0.29*  &  0.71*    &     HII   &   DS   \\
912   &  4413  &  Sbc  &  12.97 & 17 &  9.85 &  3.14   &  0.94   &  0.32   &   -       &     HII   &   LOI  \\
921   &  4412  &  Sbc  &  13.14 & 17 &  9.64 &  2.19   &  0.92   &  0.42*  &  0.31*    &     LIN   &   NED  \\
938   &  4416  &  Sc   &  13.28 & 17 &  9.74 &  2.80   &  0.77   &  0.33*  &  0.20*    &     HII   &   DS   \\
939   &     -  &  Sc   &  12.92 & 23 &  9.87 &  3.25   &  0.71   &  0.89   &   -       &   Sey/LIN &   SDSS \\
950   &     -  & Sm    &  14.49 & 17 &  8.63 &  2.78   &   -     &  0.19*  &   -       &     HII   &   DS   \\
957   &  4420  &  Sc   &  12.67 & 17 &  9.91 &  3.23   &  0.34   &  0.44   &  0.62     &  LIN/HII  &   SDSS \\
958   &  4419  & Sa    &  12.13 & 17 & 10.61 &  5.15   &  0.82   &  1.61   &   -       &     LIN   &   Ho   \\
971   &  4423  &  Sd   &  14.28 & 23 &  9.55 &  3.44   &  0.43   &  0.11*  &  1.91*    &     HII   &   DS   \\
975   &     -  &  Scd  &  13.58 & 23 &  9.66 &  3.03   &  0.61   &  0.12*  &   -       &     HII   &   DS   \\
979   &  4424  & Sa    &  12.32 & 23 & 10.38 &  3.51   &  0.83   &  0.34   &  0.20     &     HII   &   Ho   \\
980   &     -  &  Scd  &  14.17 & 17 &  8.15 &  2.07   &  0.44   &  0.14*  &  1.71*    &     HII   &   DS   \\
984   &  4425  & Sa    &  12.82 & 17 & 10.11 &  4.68   &   -     &    -    &   -       &     NEL   &   SDSS \\
1002  &  4430  &  Sc   &  12.48 & 23 & 10.22 &  2.73   &  0.81   &  0.43*  &   -       &  LIN/HII  &   DS   \\
1011  &     -  & Sdm   &  14.85 & 17 &  8.90 &  2.46   &  0.23   &    -    &   -       &     HII   &   DS   \\
1043  &  4438  & Sb    &  10.91 & 17 & 10.83 & 10.21   &  1.13   &  1.97   &   -       &     LIN   &   Ho   \\
1047  &  4440  & Sa    &  12.48 & 17 & 10.34 &  7.42   &   -     &    -    &   -       &     NEL   &   DS   \\
1060  &     -  & Sm    &  15.00 & 17 &   -   &   -     &  0.37   &  0.07   &  1.60     &     HII   &   SDSS \\
1086  &  4445  & S..   &  13.66 & 23 & 10.08 &  3.04   &  0.45   &    -    &   -       &   LIN/HII &   DS   \\
1091  &     -  &  Sbc  &  14.60 & 23 &  9.18 &  3.77   &  0.65   &  0.11*  &   2.28*   &     HII   &   DS   \\
1110  &  4450  & Sab   &  10.93 & 17 & 10.94 &  4.33   &   1.4   &  6.74   &   -       &     LIN   &   Ho   \\
1118  &  4451  &  Sc   &  13.31 & 23 & 10.08 &  2.84   &  0.23   &  0.32*  &  0.17*    &     HII   &   DS   \\
1126  &     -  &  Sc   &  13.30 & 17 & 10.06 &  3.28   &  0.69   &  0.37   &   -       &     HII   &   SDSS \\
1145  &  4457  & Sb    &  11.66 & 17 & 10.64 &  7.17   &  1.34   &  3.25   &   -       &     LIN   &   Ho   \\
1156  &     -  &  Scd  &  14.13 & 17 &   -   &   -     &  0.43   &    -    &   -       &     HII   &   SDSS \\
1158  &  4461  & Sa    &  12.09 & 17 & 10.54 &  7.45   &   -     &    -    &   -       &     NEL   &   DS   \\
1189  &     -  &  Sc   &  13.70 & 17 &  9.30 &  2.42   &  0.54   &    -    &   -       &     HII   &   DS   \\
1190  &  4469  & Sa    &  12.22 & 23 & 10.78 &  4.65   &  0.69   &  0.8    &   -       &   Sey/LIN &   SDSS \\
1193  &  4466  &  Sc   &  14.62 & 17 &  9.28 &  2.61   &  0.18   &  0.25*  & 0.71*     &    HII    &   DS   \\
1205  &  4470  &  Sc   &  13.04 & 17 &  9.73 &  2.33   &   0.1   &  0.26*  & 0.51*     &    HII    &   Ho   \\
1217  &     -  & Sm    &  14.59 & 17 &  8.94 &  2.80   &   -     &    -    &   -       &     NEL   &   DS   \\
1249  &     -  & Im    &  14.75 & 17 &  8.47 &  2.73   &   -     &    -    &   -       &     NEL   &   DS   \\
1266  &     -  & Sdm   &  14.63 & 17 &   -   &   -     &   -     &    -    &   -       &     HII   &   SDSS \\
1290  &  4480  & Sb    &  13.09 & 17 &  9.91 &  3.88   &  0.79   &  1.46   &   -       &   Sey/LIN &   SDSS \\
1326  &  4491  & Sa    &  13.41 & 17 &  9.78 &  2.87   &   -     &    -    &   -       &     HII   &   DS   \\
1330  &  4492  & Sa    &  13.17 & 17 & 10.13 &  4.41   &  1.21   &    -    &   -       &     NEL   &   LOI  \\
1375  &     -  &  Sc   &  12.00 & 17 &  9.35 &  2.78   &  0.72   &  0.27*  &  0.54*    &     HII   &   DS   \\
1379  &  4498  &  Sc   &  12.62 & 17 &  9.84 &  2.27   &  0.77   &  0.28*  &  0.28*    &     HII   &   DS   \\
1393  &     -  &  Sc   &  14.01 & 17 &  9.47 &  3.08   &  0.45   &  0.26*  &  0.31*    &     HII   &   DS   \\
1401  &  4501  &  Sbc  &  10.27 & 17 & 11.18 &  3.73   &  1.22   &  1.42   &   -       &     Sey   &   Ho   \\
1410  &  4502  & Sm    &  14.57 & 17 &  9.02 &  2.80   &  0.33   &  0.20*  &  1.05*    &     HII   &   DS   \\
1412  &  4503  & Sa    &  12.12 & 17 & 10.53 &  5.95   &   -     &    -    &   -       &     NEL   &   LOI  \\
1419  &  4506  & 18    &  13.64 & 17 &  9.66 &  3.44   &  0.57   &  0.35   &   -       &     HII   &   SDSS \\
1450  &     -  &  Sc   &  13.29 & 17 &  9.44 &  2.93   &  0.61   &  0.24*  &   0.42*   &     HII   &   DS   \\
1508  &  4519  &  Sc   &  12.34 & 17 &  9.93 &  3.17   &  0.83   &  0.22*  &   1.17*   &     HII   &   DS   \\
1516  &  4522  &  Sbc  &  12.73 & 17 &  9.85 &  3.12   &  0.46   &  0.33   &  0.22     &     HII   &   SDSS \\
1524  &  4523  &  Sd   &  13.51 & 17 &  9.49 &   -     &  0.75   &  0.14*  &  0.97*    &     HII   &   DS   \\
1532  &     -  &  Sc   &  14.05 & 17 &  9.55 &  1.69   &  0.45   &  0.25   &  0.20     &     HII   &   SDSS \\
1540  &  4527  & Sb    &  11.32 & 17 & 10.91 &  6.79   &  1.02   &  0.45*  &   -       &     LIN   &   Ho   \\
1552  &  4531  & Sa    &  12.58 & 17 & 10.22 &  3.00   &   -     &  2.69   &   -       &   Sey/LIN &   MDS  \\
1554  &  4532  & Sm    &  12.30 & 17 &  9.90 &  2.92   &  0.26   &  0.12*  &  2.00*    &     HII   &   Ho   \\
1555  &  4535  &  Sc   &  10.51 & 17 & 10.76 &  2.18   &  1.41   &  0.41   &   -       &  LIN/HII  &   LOI  \\
1557  &  4533  &  Scd  &  14.53 & 17 &  9.12 &  2.83   &  0.34   &  0.22   &  0.88     &     HII   &   SDSS \\
1562  &  4536  &  Sc   &  11.01 & 17 & 10.71 &  6.96   &  1.61   &  0.50   &   -       &   LIN/HII &   LOI  \\
1569  &     -  &  Scd  &  15.00 & 17 &  8.42 &  3.16   &  0.54   &  0.12   &  1.71     &     HII   &   SDSS \\
1575  &     -  & Sm    &  13.98 & 17 &  9.31 &  2.44   &  0.41   &  0.30*  &  0.08*    &     HII   &   DS   \\
1581  &     -  & Sm    &  14.55 & 17 &  8.74 &  2.77   &   -     &  0.05*  &  0.65*    &     HII   &   DS   \\
1588  &  4540  &  Scd  &  12.81 & 17 & 10.05 &  2.25   &  0.49   &  0.34*  &   -       &     Sey   &   NED  \\
1615  &  4548  & Sb    &  10.98 & 17 & 10.91 &  4.34   &  1.46   &  22.14  &   -       &     LIN   &   Ho   \\
1624  &  4544  &  Sc   &  13.89 & 17 &  9.68 &  2.77   &  0.51   &  0.31   &  0.42     &     HII   &   SDSS \\
1673  &  4567  &  Sc   &  12.08 & 17 & 10.41 &  2.73   &   -     &  0.39   &	-      &     HII   &   Ho   \\
1675  &     -  & Pec   &  14.47 & 17 &  8.96 &  2.87   &  0.37   &  0.17*  &   1.25*   &     HII   &   DS   \\
1676  &  4568  &  Sc   &  11.70 & 17 & 10.71 &  4.27   &   -     &  0.34*  &   0.14*   &     HII   &   Ho   \\
1678  &     -  &  Sd   &  13.70 & 17 &  8.81 &  2.91   &  0.54   &  0.10*  &   2.05*   &     HII   &   DS   \\
1686  &     -  & Sm    &  13.95 & 17 &  9.32 &  2.98   &  0.45   &  0.20*  &   1.51*   &     HII   &   DS   \\
1690  &  4569  & Sab   &  10.25 & 17 & 11.05 &  4.37   &   1.8   &  1.52   &   -       &     LIN   &   Ho   \\
1696  &  4571  &  Sc   &  11.81 & 17 & 10.38 &  2.94   &  0.76   &  0.39*  &   -       &     HII   &   DS   \\
1699  &     -  & Sm    &  14.11 & 17 &  8.83 &  2.64   &  0.57   &  0.11*  &  4.85*    &     HII   &   DS   \\
1725  &     -  &Sm/BCD &  14.51 & 17 &  8.97 &  2.92   &  0.34   &  0.14*  &  2.14*    &     HII   &   DS   \\
1726  &     -  & Sdm   &  14.54 & 17 &  8.64 &  2.73   &   -     &    -    &  1.18*    &     HII   &   DS   \\
1727  &  4579  & Sab   &  10.56 & 17 & 11.11 &  4.51   &  1.43   &  2.88   &   -       &     Sey   &   Ho   \\
1730  &  4580  &  Sc   &  12.61 & 17 & 10.22 &  2.68   &   0.6   &  1.82   &   -       &   Sey/LIN &   SDSS \\
1757  &  4584  & Sa    &  13.60 & 17 &  9.54 &  3.53   &  0.63   &  0.34   &  0.11     &     HII   &   SDSS \\
1758  &     -  &  Sc   &  14.99 & 17 &  9.03 &  3.47   &   0.3   &  0.08*  &  1.31*    &     HII   &   DS   \\
1760  &  4586  & Sa    &  12.54 & 17 & 10.32 &  6.11   &  0.78   &  1.10   &  8.34     &     Sey   &   SDSS \\
1780  &  4591  & Sb    &  13.70 & 17 &  9.63 &  3.01   &  0.43   &  0.39   &  0.20     &     HII   &   SDSS \\
1791  &     -  &Sm/BCD &  14.67 & 17 &  8.95 &  2.86   &  0.18   &    -    &   -       &     HII   &   DS   \\
1811  &  4595  &  Sc   &  12.92 & 17 &  9.85 &  2.71   &  0.61   &  0.29*  &  0.60*    &     HII   &   DS   \\
1813  &  4596  & Sa    &  11.51 & 17 & 10.84 &  5.44   &   -     &  0.67   &   -       &     LIN   &   Ho   \\
1859  &  4606  & Sa    &  12.52 & 17 & 10.10 &  3.40   &  0.82   &  0.63*  &   -       &  Sey/LIN  &   DS   \\
1868  &  4607  &  Scd  &  13.75 & 17 &  9.92 &  3.02   &  0.42   &  0.68   &  1.28     &     LIN   &   SDSS \\
1923  &  4630  &  Sbc  &  13.14 & 17 &  9.83 &  2.48   &  0.71   &  0.28   &   -       &     HII   &   LOI  \\
1929  &  4633  &  Scd  &  13.77 & 17 &  9.46 &  3.14   &  0.49   &  0.20*  &   1.05*   &    HII    &   DS   \\
1932  &  4634  &  Sc   &  13.19 & 17 &  9.99 &  3.08   &  0.15   &  0.40*  &   0.45*   &    HII    &   DS   \\
1943  &  4639  & Sb    &  12.19 & 17 & 10.26 &  3.54   &  0.94   &  1.18   &   -       &     Sey   &   Ho   \\
1955  &  4641  & S/BCD &  14.32 & 17 &  9.36 &  4.17   &  1.04   &  0.24   &  0.31     &     HII   &   SDSS \\
1972  &  4647  &  Sc   &  12.03 & 17 & 10.51 &  3.06   &   -     &  0.28   &   -       &     HII   &   Ho   \\
1987  &  4654  &  Sc   &  11.14 & 17 & 10.66 &  2.93   &  1.01   &  0.28   &  0.20     &     HII   &   Ho   \\
1999  &  4659  & Sa    &  13.08 & 17 & 10.04 &  6.03   &   -     &    -    &   -       &     NEL   &   DS   \\
2023  &     -  &  Sc   &  13.86 & 17 &  9.17 &  3.09   &  0.36   &  0.15*  &  1.91*    &    HII    &   DS   \\
2033  &     -  & BCD   &  14.65 & 17 &  8.66 &  3.70   &  0.38   &  0.11*  &  1.48*    &     HII   &   DS   \\
2058  &  4689  &  Sc   &  11.55 & 17 & 10.48 &  2.80   &  1.02   &  1.58   &   -       &   Sey/LIN &   SDSS \\
2070  &  4698  & Sa    &  11.53 & 17 & 10.75 &  5.78   &  1.22   &  4.84   &   -       &     Sey   &   Ho   \\
13046 &  4045  & Sa    &  13.50 & 17 & 10.25 &  5.05   &  0.97   &  0.89   &  1.28     &     LIN   &   SDSS \\
14063 &  4517  &  Sc   &  12.40 & 17 & 10.78 &  2.88   &  0.31   &  0.24*  &  0.85*    &     HII   &   DS   \\
14110 &  4632  &  Sc   &  12.60 & 17 & 10.13 &  4.03   &  0.57   &  0.37   &   -       &     HII   &   LOI  \\
15031 &  4771  &  Sc   &  13.30 & 17 & 10.14 &  3.73   &   0.5   &  1.30   &   -       &   Sey/LIN &   SDSS \\
15032 &  4772  & Sa    &  12.90 & 17 & 10.36 &  8.40   &  1.32   &  1.31*  &   -       &     LIN   &   Ho   \\
15037 &     -  &  Scd  &  13.60 & 17 &  9.13 &  2.12   &   0.8   &  0.36   &  0.77     &     HII   &   SDSS \\
15049 &  4845  & Sb    &  12.90 & 17 & 10.57 &  4.91   &   0.9   &  0.85   &  1.69     &     LIN   &   SDSS \\
15055 &  4904  &  Sc   &  13.20 & 17 & 10.04 &  2.70   &  0.84   &  0.32   &   -       &     HII   &   LOI  \\
41041 &  4116  &  Scd  &  13.00 & 17 &  9.87 &  2.27   &  0.89   &  0.23*  & 0.51*     &     HII   &   DS   \\
41042 &  4123  &  Sc   &  13.10 & 17 &  9.94 &  3.61   &   1.3   &  0.54   &  0.26     &   LIN/HII &   Ho   \\
43028 &  4688  &  Sc   &  14.50 & 17 &  9.52 &  2.41   &   0.7   &    -    &   -       &   LIN/HII &   Ho   \\
43034 &  4701  &  Sc   &  13.10 & 17 &  9.79 &  3.25   &  0.61   &  0.24*  &  0.88*    &     HII   &   DS   \\
43041 &  4713  &  Sc   &  12.30 & 17 &  9.97 &  2.93   &  0.64   &  0.73   &  1.42     &   LIN/HII &   Ho   \\
43054 &  4765  &  Scd  &  13.00 & 17 &  9.43 &  2.75   &  0.56   &  0.11*  &  2.60*    &    HII    &   DS   \\
43066 &  4799  &  Sd   &  14.40 & 17 &  9.76 &  4.59   &  0.63   &  1.70   &	-      &   Sey/LIN &   SDSS \\
43071 &  4808  &  Sc   &  12.50 & 17 & 10.12 &  3.32   &  0.51   &  0.41*  &  0.68*    &  LIN/HII  &   DS   \\
43093 &  4900  &  Sc   &  12.80 & 17 & 10.27 &  1.82   &  0.94   &  0.33   &  0.14     &     HII   &   SDSS \\
69036 &  4067  & Sb    &  13.20 & 17 &  9.67 &  3.34   &  0.88   &  0.79   &   -       &   Sey/LIN &   MDS  \\
71060 &  4746  &  Sd   &  13.30 & 17 &  9.87 &  2.85   &  0.16   &  0.41   &  0.40     &   LIN/HII &   SDSS \\
71068 &  4779  &  Sc   &  13.50 & 17 &  9.76 &  3.56   &  0.91   &  0.41   &   -       &  LIN/HII  &   LOI  \\
71092 &  4866  & Sa    &  11.90 & 17 & 10.54 &  6.28   &   -     &    -    &   -       &     LIN   &   Ho   \\
100004&  4651  &  Sc   &  11.30 & 17 & 10.52 &  2.98   &  0.82   &  1.13   &   -       &     LIN   &   Ho   \\
100015&  4758  &  Scd  &  14.10 & 17 &  9.77 &  3.83   &  0.42   &    -    &   -       &     HII   &   DS   \\
%\hline
\multicolumn{4}{c}{not  measured  spectroscopically} \\
%not & measured & spectroscopically  \\
34    &     -  &  Sc   &  14.65 & 32 &   -   &   -     &   -     &    -    &   -      &      -    &   -    \\
48    &     -  & Sdm   &  14.30 & 32 &  9.28 &  2.70   &  0.53   &    -    &   -      &      -    &   -    \\
99$^\dagger$   &     -  & Sa    &  14.81 & 32 &  9.56 &  3.02   &   -     &    -    &   -      &      -    &   -    \\
275   &     -  & Im    &  14.54 & 32 &  9.35 &  3.10   &   -     &    -    &   -      &      -    &   -    \\
331   &     -  & Pec   &  15.00 & 32 &   -   &   -     &   0.3   &    -    &   -      &      -    &   -    \\
362$^\dagger$  &  4266  & Sa    &  14.51 & 32 & 10.46 &  3.60   &   -     &    -    &   -      &      -    &   -    \\
434   &  4287  & S..   &  14.65 & 23 &  9.49 &  3.93   &  0.44   &    -    &   -      &      -    &   -    \\
449$^\dagger$  &  4289  &  Sbc  &  14.34 & 17 &  9.72 &  5.70   &   0.7   &    -    &   -      &      -    &   -    \\
514   &     -  &  Sc   &  14.70 & 23 &   -   &   -     &   -     &    -    &   -      &      -    &   -    \\
517   &     -  & Sab   &  14.90 & 17 &  9.42 &   -     &  0.43   &    -    &   -      &      -    &   -    \\
531   &     -  & Sa    &  15.00 & 17 &  8.79 &   -     &  0.33   &    -    &   -      &      -    &   -    \\
567$^\dagger$  &     -  &  Scd  &  14.36 & 23 &  9.16 &  2.65   &  0.62   &    -    &   -      &      -    &   -    \\
697$^\dagger$  &     -  &  Sc   &  14.17 & 23 &  9.66 &  3.02   &  0.73   &    -    &   -      &      -    &   -    \\
768   &     -  &  Sc   &  14.91 & 17 &  8.90 &  3.16   &  0.32   &    -    &   -      &      -    &   -    \\
826   &     -  & Im    &  15.00 & 17 &   -   &   -     &   -     &    -    &   -      &      -    &   -    \\
1017  &     -  & Im    &  14.50 & 23 &   -   &   -     &   -     &    -    &   -      &      -    &   -    \\
1114  &     -  & Im    &  14.82 & 17 &  8.98 &   -     &   -     &    -    &   -      &      -    &   -    \\
1435  &     -  & Im    &  14.63 & 17 &  8.95 &   -     &   -     &    -    &   -      &      -    &   -    \\
1442$^\dagger$ &     -  &  Sd   &  14.82 & 17 &   -   &   -     &  0.31   &    -    &   -      &      -    &   -    \\
1465  &     -  & Im    &  15.00 & 17 &   -   &   -     &   -     &    -    &   -      &      -    &   -    \\
1468  &     -  & Im    &  15.00 & 17 &   -   &   -     &  0.19   &    -    &   -      &      -    &   -    \\
1529  &     -  & Sdm   &  14.63 & 17 &   -   &   -     &  0.07   &    -    &   -      &      -    &   -    \\
1566  &     -  &  Sd   &  14.80 & 17 &   -   &   -     &  0.33   &    -    &   -      &      -    &   -    \\
14062$^\dagger$&     -  &  Scd  &  14.10 & 17 &  9.63 &  2.87   &  0.57   &    -    &   -      &      -    &   -    \\

\end{longtable}
\end{tiny}
Col 1: VCC or CGCG designation; Col 2: NGC name; Col 3: Hubble type from the VCC; Col 4: Photographic magnitude
from the VCC; Col 5: Distance; Col 6: H-band luminosity; Col 7: H-band light concentration index; Col 8: $r$-band
nuclearity parameter; Col 9: ratio of N[II]/H$\alpha$. An asterisk indicates Drift-Scan spectra;
Col 10: ratio of O[III]/H$\beta$. An asterisk indicates Drift-Scan spectra;  Col 11:
Adopted spectral classification; Col 12: reference to the spectral classification: Ho=from Ho et al.(1997);
SDSS: nuclear spectrum from the SDSS; DS=drift-scan spectrum; LOI=spectrum taken at Loiano; MDS= Modified drift-scan.
$^\dagger$ refers to the galaxies added in proofs (see footnote 2).
\end{document}